\documentclass[prd,10pt,aps,showpacs,superscriptaddress,nofootinbib,notitlepage,floatfix]{revtex4-1}
\usepackage{slashed}
\usepackage{epsfig}
\usepackage{eurosym}
\usepackage{slashed}
\usepackage{epsfig}
\usepackage{color}

\newcommand{\beqn}{\begin{eqnarray}}
\newcommand{\eeqn}{\end{eqnarray}}

\begin{document}

\title{The type II Weyl semimetals at low temperatures: chiral anomaly,
elastic deformations, zero sound}
\author{M.A. Zubkov }
\thanks{On leave of absence from Institute for Theoretical and Experimental
Physics, B. Cheremushkinskaya 25, Moscow, 117259, Russia}
\affiliation{Physics Department, Ariel University, Ariel 40700, Israel}

\author{M. Lewkowicz}
\affiliation{Physics Department, Ariel University, Ariel 40700, Israel}

\begin{abstract}
We consider the properties of the type II Weyl semimetals at low
temperatures basing on the particular tight - binding model. In the presence
of electric field directed along the line connecting the Weyl points of
opposite chirality the occupied states flow along this axis giving rise to
the creation of electron - hole pairs. The electrons belong to a vicinity of
one of the two type II Weyl points while the holes belong to the vicinity of
the other. This process may be considered as the manifestation of the chiral
anomaly that exists without any external magnetic field. It may be observed
experimentally through the measurement of conductivity. Next, we consider
the modification of the theory in the presence of elastic deformations. In
the domain of the considered model, where it describes the type I Weyl
semimetals the elastic deformations lead to the appearance of emergent
gravity. In the domain of the type II Weyl semimetals the form of the Fermi
surface is changed due to the elastic deformations, and its fluctuations
represent the special modes of the zero sound. We find that there is one -
to one correspondence between them and the sound waves of the elasticity
theory. Next, we discuss the influence of the elastic deformations on the
conductivity. The particularly interesting case is when our model describes
the intermediate state between the type I and the type II Weyl semimetal.
Then without the elastic deformations there are the Fermi lines instead of
the Fermi points/Fermi surface, while the DC conductivity vanishes. However,
even small elastic deformations may lead to the appearance of large
conductivity.
\end{abstract}

\pacs{}
\maketitle

\affiliation{Physics Department, Ariel University, Ariel 40700, Israel}

\affiliation{Physics Department, Ariel University, Ariel 40700, Israel}


\affiliation{Physics Department, Ariel University, Ariel 40700, Israel}

\affiliation{Physics Department, Ariel University, Ariel 40700, Israel}

\section{Introduction}

The Dirac semimetals, where the low energy physics is described by Dirac
equation, were discovered recently \cite%
{semimetal_discovery,semimetal_discovery2,semimetal_discovery3,ref:semimetal:3,ZrTe5:2,Bi2Se3,ref:semimetal:4}%
. In those materials the Fermi points corresponding to the left - handed and
the right - handed fermions coincide. The materials, where in each pair of
the left - handed and the right - handed fermions the Weyl points of
opposite chirality are separated by the finite distance in momentum space
are called Weyl semimetals and were also discovered recently \cite%
{WeylSemimetalDiscovery}.

In \cite{VZ} it was proposed that there may exist the type of the Weyl
fermions, which was later discovered in the materials called the type II
Weyl semimetals \cite{W2}. The Dirac cone of the quasiparticle dispersion is
tilted for the type II Weyl fermions in such a way, that instead of the
single Fermi point the two Fermi pockets appear touching each other at the
type II Weyl point. Surprisingly the fermions inside the horizon of the
equilibrium black holes discussed in \cite{VolovikBH} reveal the analogy to
the type II Weyl semimetals \cite{VolovikBHW2}. Even more exotic forms of
the Weyl fermions were proposed in \cite{NissinenVolovik2017a}.

The mentioned above materials are of especial importance because the physics
of the low energy fermionic excitations inside them is described by the same
equations as the elementary particles of the high energy physics. Although
certain difference appears when the interactions are taken into account,
those materials are able to simulate within laboratory the effects specific
for the high energy physics. To some extent, such materials may be used as
analogue computers for the simulation of the elementary particle physics.

Among those effects, in particular, there are the effects of chiral anomaly.
In the $3+1$ d systems with massless fermions (both in high energy physics
and in condensed matter physics) the expressions for anomalies in fermion
currents in the presence of external fields (without interaction between the
fermions) may be derived using the explicit solution of Weyl equation. The
spectral flow in the presence of gauge fields and gravity should result in
the appearance of anomalies in the particle currents \cite{Volovik2003}.
There are also the other methods developed for the calculation of anomalies
in fermion currents (Fujikawa method \cite{Weinberg}, perturbative
calculations, consideration of the $4+1$ d topological isolators with $3+1$
d chiral fermions living on their boundaries \cite{Parrikar2014}, etc). The
derivation that uses the spectral flow is in a certain sense distinguished
because it does not refer to the integration measure over fermions in the
functional integral (contrary to the Fujikawa method) and it typically does
not require any regularization (contrary to all mentioned above alternative
methods).

At the present moment the expressions for the anomalies in flat background
are well - established and commonly accepted. All mentioned above methods of
anomaly calculation give the same result. It is worth mentioning, however,
that the consideration of the case when background gravity is present is
more involved than the consideration of the case of flat background. There
are contradictions in the expressions for chiral anomaly in this case:
different methods of calculation give different expressions (see references
\cite{Zanelli,Parrikar2014,Mielke,obukhov,yajima,Obukhov:1997pz,soo}, where
those expressions are presented).

Certain observable effects in the condensed matter systems may be related to
anomalies - for example, the appearance of Kopnin force acting on vortices
in superfluid helium \cite{Volovik2003}, magneto - resistance in Weyl
semimetals \cite%
{Parrikar2014,ref:transport,semimetal_effects2,semimetal_effects3,semimetal_effects6, semimetal_effects7,semimetal_effects8,semimetal_effects10,semimetal_effects11, semimetal_effects12,semimetal_effects13,Zyuzin:2012tv,tewary,Buividovich2014, SonSpivak2012,Z2015,SonYamamoto2012,ZrTe5,ChiralAnomalySemimetal,ref:diffusion, Abramchuk:2016afc}%
, etc. The nature of the chiral anomaly is pumping of particles from vacuum.
There are also the cousins of the chiral anomaly - the effects, which are
not related directly to pumping of the particles from vacuum, but
nevertheless are related intimately to anomaly. Those are the chiral
separation effect (CSE) and the chiral vortical effect (CVE) existing both
in high energy physics and in condensed matter physics (see, for example,
\cite{Volovik2003,CVE,Buividovich2014,SonSpivak2012,SonYamamoto2012,ZrTe5}
and references therein). Actually, it was proposed to derive the expressions
for the mentioned effects directly from the expression for the anomaly (see,
for example, \cite{Zyuzin:2012tv} and references therein).

In the present paper we consider the type II Weyl semimetal in the presence
of external electric field directed along the line connecting the Weyl
points of opposite chirality. It appears, that this situation is similar to
the one of the type I Weyl semimetal in the presence of magnetic field
parallel to electric field. The electric field pumps the pairs of electrons
and holes from vacuum. Electrons and holes belong to the vicinities of the
Weyl points of opposite chiralities. Therefore, we may refer to this
phenomenon as to the kind of chiral anomaly specific for the type II Weyl
semimetals, which exists without external magnetic field. It may be observed
experimentally in the clean enough materials at the sufficiently small
temperatures, when the corresponding resistivity exceeds the resistivity
existing due to impurities. It is worth mentioning, that in order to observe
the mentioned above phenomena we need that the given material does not
undergo the phase transition to the superconducting state at the considered
values of temperature. For the discussion of the latter possibility see \cite%
{TypeIIWeylSuper}.

Another important class of effects appears due to the elastic deformations
\cite{ref:LL}. In graphene \cite%
{vozmediano2,vozmediano4,vozmediano5,vozmediano6,Oliva,VZ2013,VolovikZubkov2014,Volovik:2014kja,Khaidukov:2016yfi}%
, in the type I Weyl semimetals and in the Dirac semimetals the elastic
deformations lead to the appearance of the emergent gauge field and emergent
gravitational field (see, for example, \cite%
{Horava2005,VZ2014NPB,WeylTightBinding,VZ2014NPB,Cortijo:2015jja,Chernodub:2015wxa,Zubkov:2015cba}%
). The physics of the interior of the type I Weyl semimetals describes
qualitatively the elementary particles of the high energy physics. The
elastic deformations allow to simulate elementary particles in the presence
of gravity and the gauge field. The latter appears as the variation of the
Fermi point position in momentum space while the former is the variation of
the slope of the dependence of energy on momentum.

At the same time as it was mentioned above, the type II Weyl semimetals
simulate the interior of the black holes if they undergo the transition to
the equilibrium state \cite{Z2018}. Therefore, we expect, that the elastic
deformations for the type II Weyl semimetals allow to simulate the physics
of elementary particles in curved space inside the equilibrium black holes.
Contrary to the type II Weyl semimetals the variation of the Fermi surface
does not lead to the emergent gauge field. This variation influences the
ensemble of fermionic quasiparticles in the very special way. In particular,
in the case when the elastic deformations oscillate, the corresponding
variation of the Fermi surface becomes the special mode of the zero sound.
It is the oscillation of the shape of the Fermi surface that is able to
propagate inside the material \cite{Volovik2003}. We find, that this
propagation occurs with the velocity of the sound waves of the elasticity
theory. Moreover, there is the one - to one correspondence between those
waves and the mentioned zero sound modes. This is necessary to distinguish
those zero sound modes from the conventional zero sound. The latter is
purely non - equilibrium phenomenon and is not related to any excitations of
atoms. The modes of the zero sound investigated here may be considered as
the oscillations of the Fermi surface forced by the oscillations of the
atoms of the crystal lattice.

In addition, we consider the effect of the elastic deformations on the
conductivity of clean enough type II Weyl materials at small temperatures
when the corresponding resistivity dominates over the one caused by the
impurities. (For the description of some real Weyl materials see \cite%
{conductivity,crystal_size,crystal_size2}.) In general case the contribution
to the resistivity due to the elastic deformations is linear in the
deformation tensor. Besides, we consider the special case when the
unstrained material is in the intermediate state between the type I and the
type II Weyl semimetals. In this state the Fermi lines appear. For the
discussion of the cases, when the quantum state appears with the Fermi lines
instead of the Fermi points/Fermi surface see \cite{Z2017L} and references
therein. In this marginal case the DC conductivity is absent. However, even
small elastic deformations are able to drive the material to the type II
state with the huge conductivity.

The paper is organized as follows. In Section \ref{sectHorava} we remind the
reader the general description of the emergent Weyl spinors in the
multifermion systems. In Section \ref{SectII} we explain when and how the
type II Weyl fermions appear. In Section \ref{SectEGD} we remind the reader
the description of the chiral anomaly in the type I Weyl semimetal. In
Section \ref{AnomalyII} we discuss the chiral anomaly in the particular
model of the type II Weyl semimetal that may exist without any magnetic
field. In Section \ref{SectCondII} we demonstrate how this anomaly may be
observed experimentally through the resistivity of the sufficiently clean
materials at very small temperatures. In Section \ref{SectGravI} we consider
the influence of elastic deformations on the considered toy model of the
type I/type II Weyl semimetal and derive the emergent gravitational field
for the case when the system is in the type I state. In Section \ref%
{SectElasticII} we consider the same model in the type II domain and
demonstrate how the modification of the Fermi surface appears due to the
elastic deformations. In Section \ref{SectZeroSoundPhonons} we establish
relation between the sound waves of the elasticity theory and the zero sound
modes. In Section \ref{SectIIelastic} we consider the influence of the
elastic deformation on the conductivity. In particular we discuss the
marginal case when without the deformation there are the Fermi lines, so
that the system is between the type I and the type II states. In Section \ref%
{SectConcl} we end with the conclusions.

\section{Emergent Weyl spinors in the system with the multi - component
fermions}

\label{sectHorava}

In \cite{Horava2005} it was proposed that the emergent low dimensional
fermions in a vicinity of the topologically protected Fermi - points in
general, rather complicated, fermion systems are necessarily described by
the two component Weyl spinors. The proof of this conjecture may be found,
for example, in \cite{VZ}. Below we briefly remind the reader the
corresponding constructions.

Let us consider the $n$ - component spinors $\psi $. The partition function
has the form:
\begin{equation}
Z=\int D\psi D\bar{\psi}{\rm exp}\left( i\int dt\sum_{\mathbf{x}}\bar{\psi%
}_{\mathbf{x}}(t)(i\partial _{t}-\hat{H})\psi _{\mathbf{x}}(t)\right) ,
\label{ZH}
\end{equation}%
where the Hermitian Hamiltonian $H$ is the function of momentum operator $%
\hat{\mathbf{p}}=-i\nabla $. We use here the symbol of the summation over
the points of coordinate space because first of all we have in mind the
tight - binding models of solids. However, the whole formalism works also
for the general fermion systems, in which case this symbol is to be
understood as the integral over $d^{3}x$. The key ingredient of the
construction discussed here is the repulsion between the energy levels of $%
\hat{H}$. Any small perturbation pushes apart the two crossed branches.
Therefore, only the minimal number of branches may cross each other.

The minimal number of the crossed branches at the given point is two. Near
the crossing point the physics is described by the reduced $2\times 2$
Hamiltonian and the reduced $2$ - component spinors $\Psi $ (assuming that
the Fermi energy is tuned accordingly). As a result, at low energies we may
deal with the theory that has the following partition function:
\begin{equation}
Z=\int D\Psi D\bar{\Psi}{\rm exp}\left( i\int dt\sum_{\mathbf{x}}\bar{\Psi%
}_{\mathbf{x}}(t)(i\partial _{t}-m_{k}^{L}(\hat{\mathbf{p}})\hat{\sigma}%
^{k}-m(\hat{\mathbf{p}}))\Psi _{\mathbf{x}}(t)\right)  \label{ZH__0}
\end{equation}%
where functions $m_{k}^{L},m$ are real - valued. In the presence of the CP
symmetry generated by $\mathcal{CP}=-i\sigma ^{2}$ and followed by the
change $\mathbf{x}\rightarrow -\mathbf{x}$ ($\mathcal{CP}\Psi (\mathbf{x}%
)=-i\sigma ^{2}\bar{\Psi}^{T}(-\mathbf{x})$) the term with $m(\mathbf{p})$
is forbidden while the Hamiltonian $\hat{H}$ can be represented as
\[
\hat{H}=\sum_{k=1,2,3}m_{k}^{L}(\mathbf{p})\hat{\sigma}^{k}
\]%
It is worth mentioning that because of the CPT theorem actually the CP
symmetry coincides with the time reversal symmetry. Therefore, in condensed
matter physics the given situation is typically referred to as the time
reversal invariant.

The above mentioned nontrivial topology means that the topological invariant
\begin{equation}
N=\frac{e_{ijk}}{8\pi }~\int_{\sigma }dS^{i}~\hat{m}^{L}\cdot \left( \frac{%
\partial \hat{m}^{L}}{\partial p_{j}}\times \frac{\partial \hat{m}^{L}}{%
\partial p_{k}}\right) ,\quad \hat{m}^{L}=\frac{{m}^{L}}{|{m}^{L}|}
\label{NH}
\end{equation}%
is nonzero, where $\sigma $ is the $S^{2}$ surface around the crossing point
of the two branches. If $N=1$ in Eq.(\ref{NH}), then the expansion near the
branches crossing point $p_{j}^{(0)}$ gives
\begin{equation}
m_{i}^{L}(\mathbf{p})=f_{i}^{j}(p_{j}-p_{j}^{(0)})\,.
\label{A(K)-expansion0H}
\end{equation}%
Here by $f_{i}^{j}$ we denote the coefficients in the expansion. Thus we
come to
\begin{equation}
Z=\int D\Psi D\bar{\Psi}\mathrm{exp}\left( i\int dt\sum_{\mathbf{x}}\bar{\Psi%
}_{\mathbf{x}}(t)(i\partial _{t}-f_{k}^{j}(\hat{\mathbf{p}}_{j}-p_{j}^{(0)})%
\hat{\sigma}^{k})\Psi _{\mathbf{x}}(t)\right)  \label{ZHH}
\end{equation}

Now let us suppose, that a weak inhomogeneity is present in the given
system. It may be caused, for example, by the elastic deformations, or by
other reasons. Anyway, in a vicinity of each point within the material the
partition function is described by Eq. (\ref{ZHH}), but the coefficients in
this expression as well as the position of the branch crossing point vary.
This situation has been discussed in some details for graphene \cite%
{vozmediano2,vozmediano4,vozmediano5,vozmediano6,Oliva,VZ2013,VolovikZubkov2014,Volovik:2014kja}%
. and also for the three - dimensional materials called Dirac and Weyl
semimetals \cite{Z2015,VZ2014NPB,Chernodub:2015wxa,Cortijo:2015jja}. Near
the branches crossing point we obtain the following fermionic action
\begin{equation}
S_{R}=\frac{1}{2}\int d^{4}x|\mathbf{e}|[\bar{\Psi}ie_{b}^{j}(x){\sigma }^{b}%
\mathcal{D}_{j}\Psi -[\mathcal{D}_{j}\bar{\Psi}]ie_{b}^{j}(x){\sigma }%
^{b}\Psi ]  \label{SHe_3sW2}
\end{equation}%
Here
\begin{equation}
i\mathcal{D}_{\mu }=i\partial _{\mu }+A_{\mu }(x)
\end{equation}%
is the covariant derivative. It corresponds to the $U(1)$ gauge field $%
A_{\mu }$, while $A_{\mu }$ represents the position - dependent Fermi point
(that is the emergent gauge field). The field $e_{a}^{j}$ may be interpreted
as the emergent vierbein field, $\eta _{ab}$ is metric of Minkowski space.
Internal $SO(3,1)$ indices are denoted by Latin letters $a,b,c,...$ while
the space - time indices are denoted by Greek letters or Latin letters $%
i,j,k,...$ The inverse vierbein is denoted by $e_{i}^{a}$ (it is assumed,
that $e_{a}^{i}e_{j}^{a}=\delta _{j}^{i}$). The determinant of $e_{i}^{a}$
is denoted by $|\mathbf{e}|$. By $\sigma ^{a}$ we denote Pauli matrices, and
imply $\sigma ^{0}=1$.

Without elastic deformations the emergent gauge field $A_{\mu }$ is the same
for both left - handed and right - handed fermions, does not depend on
coordinates, and its space - components are given by the position of the
unperturbed Dirac point $\mathbf{K}^{(0)}$ (the time component of $\mathbf{A}
$ vanishes in this case). The vierbein in the absence of elastic
deformations is given by
\begin{equation}
|e^{(0)}|=v_{F},\quad e_{a}^{(0)i}=\hat{f}_{a}^{i},\quad
e_{0}^{(0)i}=0,\quad e_{a}^{(0)0}=0,\ \ \ \ e_{0}^{(0)0}=\frac{1}{v_{F}}
\label{Connection00}
\end{equation}%
where $a,i,j,k=1,2,3$, and $v_{F}\hat{f}_{a}^{i}$ is the anisotropic Fermi
velocity.

Thus Eq. (\ref{SHe_3sW2}) may be considered as the action for the right -
handed Weyl fermions. They correspond to the topological invariant of Eq. (%
\ref{NH}) $N=1$. For $N=-1$ we come to the action for the emergent left -
handed Weyl fermions:
\begin{equation}
S_L = \frac{1}{2} \int d^4x |\mathbf{e}| [\bar{\Psi} i e_b^j(x) \bar{\sigma}%
^b \mathcal{D}_j \Psi - [\mathcal{D}_j\bar{\Psi}] i e_b^j(x) \bar{\sigma}^b
\Psi ]  \label{SHe_3sW6}
\end{equation}
Here $\bar{\sigma}^0 = 1$, $\bar{\sigma}^a = - \sigma^a$ for $a=1,2,3$ while
\begin{equation}
i\mathcal{D}_\mu = i\partial_\mu + A_\mu(x)  \label{DA}
\end{equation}

The solids, where the above discussed emergent Weyl fermions appear are
called the \textit{type I Weyl semimetals}. According to the Nielsen -
Ninomiya theorem the number of the emergent left handed fermions precisely
coincides with the number of the right - handed ones. The materials, in
which the positions of the right - handed and the left - handed Weyl
fermions coincide are called \textit{Dirac semimetals}. The typical Weyl and
Dirac semimetals \cite%
{semimetal_discovery,semimetal_discovery2,semimetal_discovery3} are
anisotropic, that results in the anisotropic Fermi velocity corresponding to
$3\times 3$ matrix $\hat{f} = \mathrm{diag}(\nu^{-1/3},\nu^{-1/3},\nu^{2/3})$
with $\nu \ne 1$. Notice, that in the presence of elastic deformations, in
principle, the vielbeins (as well as the emergent gauge fields) may differ
for the (left - handed or the right - handed) fermions incident at the
different Fermi points.

\section{The type II Weyl fermions}

\label{SectII}

In the present section we briefly recall the discussion of \cite{VZ} of the
new type of the \textit{Weyl fermions,} called in \cite{W2} the \textit{type
II}. In the absence of the mentioned above $CP$ symmetry (or, equivalently,
in the absence of the time reversal symmetry) we have the nonzero function $%
m(\mathbf{p})$ that is to be expanded around the branch crossing point $%
\mathbf{p}^{(0)}$: $m(\mathbf{p})\approx f_{0}^{j}(p_{j}-p_{j}^{(0)})$, $%
i,j=1,2,3$. Here $f_{0}^{j}$ are the new coefficients (which in the case of
elastic deformations will become space - depending). We come to the
expression for the partition function of Eq. (\ref{ZHH}), where the sum is
over $a=0,1,2,3$, and $k=1,2,3$ while $\sigma ^{0}=1$. The type II Weyl
point appears if $\Vert \mathcal{G}^{k}\Vert =\Vert
f_{0}^{j}[h^{-1}]_{j}^{k}\Vert >1$. (Here $h_{i}^{j}=f_{i}^{j}$ for $%
i,j=1,2,3$) Then there is the conical Fermi - surface given by the equation
\begin{equation}
\mathrm{Det}\, \sigma ^{a}f_{a}^{k}(p_{k}-p_{k}^{(0)})=0,\quad
k=1,2,3;\,a=0,1,2,3  \label{cfs}
\end{equation}%
That is
\begin{equation}
\pm \sqrt{%
\sum_{k=1,2,3}f_{k}^{j}f_{k}^{n}(p_{j}-p_{j}^{(0)})(p_{n}-p_{n}^{(0)})}%
+f_{0}^{j}(p_{j}-p_{j}^{(0)})=0,\quad j,k=1,2,3;  \label{cfs2}
\end{equation}%
Introducing the variable $\mathcal{P}_{j}=f_{j}^{k}(p_{k}-p_{k}^{(0)})$ we
come to the equation
\begin{equation}
\pm \Vert \mathcal{P}\Vert +\mathcal{G}^{j}\mathcal{P}_{j}=0,\quad j=1,2,3;
\label{cfs3}
\end{equation}%
One can see, that this equation indeed has the nontrivial solution of the
form of the Fermi surface for $\Vert \mathcal{G}\Vert >1$. This is the case
of the type II Weyl Fermi point. Notice, that at $\Vert
f_{0}^{j}[h^{-1}]_{j}^{a}\Vert =1$ instead of the Fermi point or the Fermi
surface there is the Fermi line. There may also be the other marginal case,
when $\mathrm{det}\,f_{a}^{k}=0$. Then the theory may become effectively $%
2+1 $ dimensional or even $1+1$ dimensional, i.e. there is the reference
frame, in which the Hamiltonian does not depend on one or even two
coordinates.

\begin{figure}[h]
\centering
\includegraphics[width=6cm]{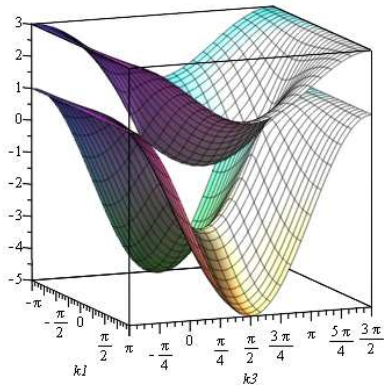}
\caption{The dispersion of the quasiparticles (in the units of $1/a$) in the
model with the Hamiltonian of Eq. (\protect\ref{WST}) at $v_f=1, v = 2$.}
\label{fig.1}
\end{figure}

For definiteness we may consider the following particular tight - binding
lattice model that describes the type II Weyl fermions:
\begin{equation}
H=\frac{v_{F}}{a}\sum_{k=1,2}\sigma ^{k}\mathrm{sin}\,(p_{k}a)+\frac{v_{F}}{a%
}\sigma ^{3}\Big(\mathrm{sin}\,(p_{3}a)+\sum_{k=1,2}(1-\mathrm{cos}%
\,(p_{k}a))\Big)-\frac{v}{a}\, \mathrm{sin}\,({p_{3}a})  \label{WST}
\end{equation}%
Here $a$ is the lattice spacing, i.e. the distance between the adjacent
sites of the tight - binding model while $v_{F}$ and $v$ are the parameters.
The first one has the meaning of the Fermi velocity. The corresponding
branches of spectrum of this Hamiltonian are given by
\begin{equation}
\mathcal{E}=\pm \frac{v_{F}}{a}\sqrt{\sum_{k=1,2}\mathrm{sin}^{2}\,(p_{k}a)+%
\Big(\mathrm{sin}\,(p_{3}a)+\sum_{k=1,2}(1-\mathrm{cos}\,(p_{k}a))\Big)^{2}}-%
\frac{v}{a}\, \mathrm{sin}\,{(p_{3}a)}
\end{equation}%
For $v<v_{F}$ there are two Fermi points of opposite chiralities
\[
K^{+}=(0,0,0),\quad K^{-}=(0,0,\pi /a)
\]%
The Hamiltonians near those two points are:
\begin{equation}
H_{+}=v_{F}\sum_{k=1,2,3}\sigma ^{k}\,p_{k}-v\,{p_{3}},\quad
H_{-}=v_{F}\sum_{k=1,2}\sigma ^{k}\,(p_{k}-K_{k}^{-})-v_{F}\sigma
^{3}(p_{3}-K_{3}^{-})+v\,(p_{3}-K_{3}^{-})
\end{equation}%
At $v>v_{F}$ we come to the description of the type II Weyl semimetal with
the Fermi pockets incident at the points $K^{\pm }$. As an illustration we
represent in Fig. \ref{fig.1} the dispersion of the quasiparticles in this
model for the particular case $v_{F}=1,v=2$.

\section{Chiral anomaly in the type I Weyl semimetals in the presence of
external magnetic field}

\label{SectEGD}

Let us start from the consideration of one particular Weyl fermion (either
right - handed or left - handed). We consider the particular case, when the
fields do not depend on the $z$ coordinate. The one-particle Hamiltonian in
the presence of the electromagnetic field $A_k$ for the right - handed
fermions is given by
\begin{eqnarray}
\mathcal{H}^{(R)} &=& {f}_a^k{\sigma}^a (\hat p_k - A_k),  \label{eq:HR}
\end{eqnarray}
while for the left - handed fermions:
\begin{eqnarray}
\mathcal{H}^{(L)} &=& - {f}_a^k{\sigma}^a (\hat p_k - A_k),  \label{eq:HR}
\end{eqnarray}
where
\begin{eqnarray}
{f}_a^k = \frac{ e_a^k}{ e_0^0}= v_F \hat{f}_a^{k}, \quad a,b,k = 1,2,3\,.
\label{eq:f:ak}
\end{eqnarray}
Let us consider the case, when the electromagnetic potential $A$ corresponds
to the constant magnetic field $B$ directed along the z axis. We suppose
that the magnetic field is nonzero within the cylinder of finite radius, and
denote the magnetic flux through the surface $S_\bot$ orthogonal to the
cylinder axis
\begin{eqnarray}
\hat{\Phi} & = & \frac{S_\bot}{2\pi} B^3,
\end{eqnarray}
The lowest Landau level modes correspond to the branches of spectrum with
the dispersion
\begin{equation}
\mathcal{E}^{(R)} \approx v_F \nu^{2/3} \mathrm{sign}(\hat{\Phi}) \, p_3\,.
\label{ER0}
\end{equation}
and
\begin{equation}
\mathcal{E}^{(L)} \approx -v_F \nu^{2/3} \mathrm{sign}(\hat{\Phi}) \, p_3\,.
\label{ER0}
\end{equation}

\begin{figure}[h]
\centering
\includegraphics[width=6cm]{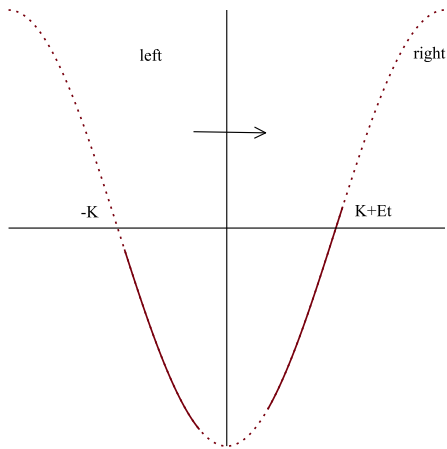}
\caption{Schematic pattern of the lowest Landau level in the presence of
external electric field $E$ and magnetic field. We represent the given
branches of spectrum as functions of $p_3$, where $p_3$ is the third
component of the total momentum. It is supposed, that the third axis is
directed along $\mathbf{K}$. Crossing of the branches occurs at the two Weyl
points of opposite chirality $\pm |\mathbf{K}|$. The left - handed branch at
$-\mathbf{K}$ is at the same time the right - handed branch at $\mathbf{K}$.
(The lines are closed through the boundary of the Brillouin zone.) The
values of energy carried by the occupied states at $t\ne 0$ are shifted by $%
Et$ (solid lines). }
\label{fig1}
\end{figure}

In the presence of an external electric field ${\bf E}$ directed
along the $z$ axis the states that correspond to those branches flow in the
correspondence with the following equation:
\begin{equation}
\langle \dot{p}_{3}\rangle =E_{3}\,.  \label{flow}
\end{equation}%
These branches of spectrum have the definite value of the spin projection to
axis $z$: $s=\frac{1}{2}\mathrm{sign}\, \hat{\Phi}$ and are given,
respectively, by the following dispersion relation (see, for example, \cite%
{Z2015,Chernodub:2015wxa} and references therein):
\[
\mathcal{E}_{R/L}(p_{3})=\pm 2sv_{F}\nu ^{2/3}p_{3}\,.
\]%
its degeneracy is:
\begin{equation}
N_{R/L}=[|\hat{\Phi}|]+1
\end{equation}%
Let us consider for definiteness the right - handed fermions. Then in vacuum
the states with $p_{z}\mathrm{sign}\, \hat{\Phi}(p_{z})<0$ are occupied.
Those states flow according to Eq. (\ref{flow}). As a result the right -
handed quasiparticles within the unit distance along the $z$ axis are pumped
from vacuum with the rate
\begin{equation}
\dot{\rho}_{R}=\frac{{E}_{z}}{2\pi S_{\bot }}\mathrm{sign}\, \hat{\Phi}%
N_{R/L},  \label{Aq}
\end{equation}%
(Here $\rho _{R}$ is the density of the right - handed quasiparticles while $%
S_{\bot }$ is the area of the sample in the $xy$ plane.) Correspondingly,
the left handed quasiparticles are pumped from vacuum with the rate
\begin{equation}
\dot{\rho}_{L}=-\frac{{E}_{z}}{2\pi S_{\bot }}\mathrm{sign}\, \hat{\Phi}%
N_{R/L},  \label{AqL}
\end{equation}%
This pattern is illustrated by Fig. \ref{fig1}. For the case $N_{R/L}\gg 1$
the conventional expression for the anomaly appears:
\[
\dot{\rho}_{R/L}=\pm \frac{1}{4\pi ^{2}}\mathbf{E}\mathbf{B}
\]%
It results in the appearance of charge carriers that consist of pairs of
left - handed electrons and right - handed holes or vice versa \cite%
{Z2015,Chernodub:2015wxa}. Those quasiparticles give the contribution to
conductivity that is not caused by the thermal excitations (see, for
example, \cite%
{ZrTe5,ref:diffusion,ref:transport,ChiralAnomalySemimetal,Abramchuk:2016afc}%
).

\section{Chiral anomaly in the type II Weyl semimetals without external
magnetic field}

\label{AnomalyII}

In this section we consider for definiteness the type II Weyl semimetal with
the Hamiltonian of Eq. (\ref{WST}). In the presence of the electric field $%
E_{z}$ directed along the line connecting the Weyl points the states flow
according to
\begin{equation}
\langle \dot{p}_{3}\rangle =E_{3}\,.  \label{flow2}
\end{equation}%
For $v<1$ one of the Weyl points is left - handed while another one is the
right - handed. One may extend this assignment of chirality to the case of
the type II Weyl points. However, in the presence of the Fermi sphere the
relevant degrees of freedom may belong to the region of momentum space
distant from the Weyl points. We extend the notion of chirality to those
states as well. For the simple model with only two Weyl points $K^{\pm }$
the state with momentum $p$ is the left - handed if $\Vert K^{-}-p\Vert
<\Vert K^{+}-p\Vert $ and it is right - handed if $\Vert K^{-}-p\Vert >\Vert
K^{+}-p\Vert $. Instead of Eqs. (\ref{Aq}) and (\ref{AqL}) we have:
\begin{equation}
\dot{\rho}_{R}=\frac{{E}_{z}}{2\pi S_{\bot }}N_{R},  \label{Aq2}
\end{equation}%
and
\begin{equation}
\dot{\rho}_{L}=-\frac{{E}_{z}}{2\pi S_{\bot }}N_{L},  \label{AqL2}
\end{equation}%
Here
\[
N_{R/L}=\frac{S_{\bot }}{2}\int \frac{dp_{x}dp_{y}}{(2\pi )^{2}}
\]%
is the integral over the whole Fermi surface.

The process of the pair creation in the presence of electric field is
counterbalanced in practise by the annihilation of the electrons and holes.
Assuming that the relaxation time is equal to $\tau$, we may estimate the
number of pairs
\[
\rho _{5}=\frac{{E}_{z} }{4\pi }\int \frac{dp_{x}dp_{y}}{(2\pi )^{2}}\tau
\]%
The Fermi surface (see Fig. \ref{fig3}) consists of the two closed pieces
having two common points (that are the type II Weyl points). Those pieces
have correspondingly the spherical topology and the topology of the Riemann
surface with two holes. In the integral of the above equation it is assumed
that the elementary area $dp_x dp_y$ is always positive.

\section{Conductivity of sufficiently clean type II Weyl semimetals at very
small temperature}

\label{SectCondII}

Assuming that the given type II semimetal does not become the
superconductor, we may estimate roughly the electric conductivity caused by
the given process at $T=0$ modeling the considered steady state by the
collection of the quasiparticles that occupy the states from the position of
the Fermi surface up to the extent of the third component of momentum $%
\Delta p_{z}=E_{z}\tau $. We may simply assume that for some reason for any
point $(p_{x},p_{y},p_{z})$ on the Fermi surface, the states between $%
(p_{x},p_{y},p_{z})$ and $(p_{x},p_{y},p_{z}+E_{z}\tau )$ are occupied.
Since $\partial \mathcal{E}/\partial p_{z}$ is the velocity of the
quasiparticle, this gives the electric current
\begin{equation}
j_{z}=N_{D}\frac{{E}_{z}}{2\pi }\int \frac{dp_{x}dp_{y}}{(2\pi )^{2}}\Big|%
\frac{\partial \mathcal{E}}{\partial p_{z}}\Big|\tau  \label{jE2}
\end{equation}%
Here $N_{D}=1$ is the number of the spectrum branches that participate in
the process at any particular point in momentum space. Here the integral is
over the Fermi surface (which has the form presented in Fig. \ref{fig3}).
Here
\[
\frac{\partial \mathcal{E}}{\partial p_{3}}={v_{F}}\, \mathrm{\cos }\,({p}%
_{3}a)\left( \pm \frac{\left[ \, \mathrm{sin}\,({p}_{3}a)+\sum_{k=1,2}(1-%
\mathrm{cos}\,(p_{k}a))\right] }{\sqrt{\sum_{k=1,2}\, \mathrm{sin}\,^{2}({p}%
_{k}a)+\left[ \mathrm{sin}\,({p}_{3}a)+\sum_{k=1,2}(1-\mathrm{cos}(p_{k}a)\,)%
\right] ^{2}}}-\frac{v}{v_{F}}\right)
\]%
The Fermi surface is the solution of the following equation:
\begin{equation}
0=\pm \frac{v_{F}}{a}\sqrt{\sum_{k=1,2}\mathrm{sin}^{2}\,(p_{k}a)+\Big(%
\mathrm{sin}\,(p_{3}a)+\sum_{k=1,2}(1-\mathrm{cos}\,(p_{k}a))\Big)^{2}}-%
\frac{v}{a}\, \mathrm{sin}\,{(p_{3}a)}
\end{equation}%
It gives
\[
\frac{\partial \mathcal{E}}{\partial p_{3}}=\, \frac{v_{F}^{2}}{v}\mathrm{%
\cos }\,({p}_{3}a)\left( \frac{\left[ \, \sum_{k=1,2}(1-\mathrm{cos}%
\,(p_{k}a))\right] }{\, \mathrm{sin}\,{(p_{3}a)}}+1-\frac{v^{2}}{v_{F}^{2}}%
\right)
\]%
%
%

Notice that the product of the relaxation time $\tau $ and the Fermi
velocity $\Big|\frac{\partial \mathcal{E}}{\partial \vec{p}}\Big|\sim v_{F}$
on the Fermi surface may be considered as the mean free path $l$ of the
quasiparticle. In turn, it may be estimated as
\[
l\sim v_{F}\tau \approx \frac{1}{\rho _{5}\sigma _{t}}
\]%
where $\sigma _{t}$ is the average transport cross - section of the electron
- hole annihilation. This gives
\[
\tau \approx \Big[\sigma _{t}v_{F}\frac{{E}_{z}}{4\pi }\int \frac{%
dp_{x}dp_{y}}{(2\pi )^{2}}\Big]^{-1/2}
\]%
The expression for the electric current receives the form
\begin{equation}
j_{z}=\sqrt{\frac{2E_{z}}{v_{F}\sigma _{t}}}\frac{\int \frac{dp_{x}dp_{y}}{%
(2\pi )^{3}}\Big|\frac{\partial \mathcal{E}}{\partial p_{z}}\Big|}{\Big[\int
\frac{dp_{x}dp_{y}}{(2\pi )^{3}}\Big]^{1/2}}  \label{jE2}
\end{equation}

\begin{figure}[h]
\centering
\includegraphics[width=6cm]{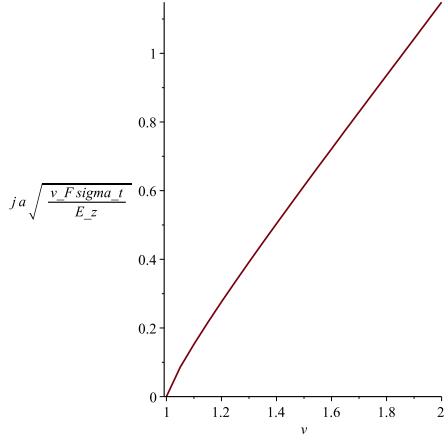}
\caption{The ratio $\frac{ja}{\protect\sqrt{v_{F}E_{z}/\protect\sigma _{t}}}$
as a function of $v$ for the model with the Hamiltonian of Eq. (\protect\ref%
{WST}).}
\label{fig.5}
\end{figure}

\begin{figure}[h]
\centering
\includegraphics[width=6cm]{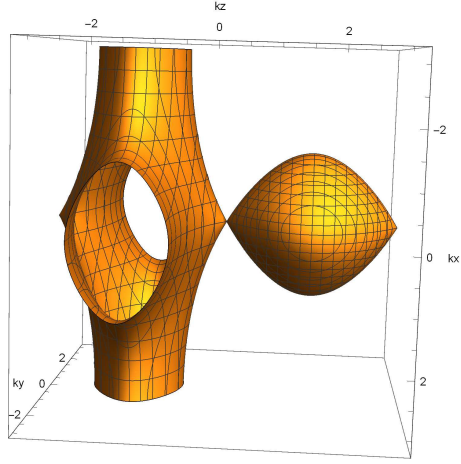}
\caption{The typical form of the Fermi surface for the type II Weyl
semimetal described by Eq. (\protect\ref{EPS}) (elastic deformations
neglected). We chose here $v_F = 1, v=2$. Here the axes correspond to the
values $k_x = p_x a, k_y = p_y a, k_z = p_z a$.}
\label{fig3}
\end{figure}

The integrals of Eq. (\ref{jE2}) have been calculated numerically for the
case $v_{F}=1$ and the result is represented in Fig. \ref{fig.5}. One can
see, that the current vanishes when the system drops from the type II phase
to the type I phase.

The typical value of the lattice spacing (interatomic distance) is $a\sim
0.25\,nm\sim 10^{-3}eV^{-1}$. Let us suppose, that the typical electric
field strength (in usual units) is of the order of $1$ $V/cm$. In order to
obtain the value of $E_{z}$ we need to transfer this value into relativistic
units. The length unit $cm$ should be expressed in $eV^{-1}$ based on the
relation $[200$ MeV$]^{-1}\approx 1$ $fm$ $=10^{-13}$ $cm,$ i.e. $1$ $cm$
corresponds to $5\times 10^{4}\,eV^{-1}$. At the same time $1$ $V$
transforms to $1$ $eV$. The resulting value of $E_{z}$ is
\begin{equation}
E_{z}\approx 2\times 10^{-5}eV^{2}
\end{equation}%
The typical value of $v_{F}$ is $\sim 1/100$. The cross - section $\sigma
_{t}$ is given by the inverse effective mass squared of the quasiparticle.
The effective mass of the quasiparticle near the Fermi surface is $m^{\ast
}\sim |\mathbf{p}/v_{F}|\sim \frac{1}{av_{F}}$. Therefore, $\sigma _{t}\sim %
\Big(\frac{\alpha }{v_{F}m^{\ast }}\Big)^{2}\sim a^{2}$ (here the effective
coupling constant $\alpha $ is of the order of unity). As a result at $%
E_{z}=1$ $V/cm$ the conductivity $\partial j/\partial E_{z}$ is
\begin{equation}
\sigma \sim \sqrt{\frac{v_{F}}{a^{2}E_{z}\sigma _{t}}}\sim 10^{8}\,eV
\label{s1}
\end{equation}%
The typical inter - collision time is
\[
\tau \sim \sqrt{\frac{a^{2}}{v_{F}E_{z}\sigma _{t}}}\sim 10^{3}\,eV^{-1}\sim
10^{-12}\,s\sim 10^{-2}\,cm/c
\]%
and the corresponding mean free path is $l\sim 10^{-4}$ cm $\sim 1000\mu $m.

The typical value of the conventional conductivity in semimetals at room
temperature \cite{conductivity} is $\sigma ^{(0)}\sim 3\times
10^{4}/(Ohm\times cm)$ $=$ $3\times 10^{4}$ $A/(V\times cm)$. One Ampere
corresponds to the flow of $\frac{1}{1.6\times 10^{-19}}\sim 6\times
10^{18}\ $electric charges. One second corresponds to the distance of $%
3\times 10^{10}$ $cm$, that is $\sim 1.5\times 10^{15}eV^{-1}$. Recall that
we absorb the elementary charge into the definition of electric field, and
the electric current is defined as the current of the electrons rather than
the current of electric charge, that is our current is equal to the
conventional electric current divided by the charge of electron. Thus in
relativistic units we have
\begin{equation}
\sigma ^{(0)}\sim 2\times 10^{3}eV
\end{equation}%
One can see, that the calculated above conductivity is
\begin{equation}
\sigma \sim 10^{5}\sigma ^{(0)}\sqrt{\frac{1\,V/cm}{E_{z}}}\sim 3\times
\sqrt{\frac{1\,V/cm}{E_{z}}}\times 10^{9}/(Ohm\times cm)  \label{s2}
\end{equation}%
This value of the conductivity is of the order of the known experimental
values of the conductivity for certain real metals at temperatures of the
order of several Kelvin. The values of the conductivity for clean enough Au
and Cu at $T\sim 1$ K may be of the same order of magnitude. The above value
remains finite and is smaller than the formally infinite value predicted by
the kinetic theory that takes into account the interaction of the
quasiparticles with phonons. For example, the version of the theory
presented in the classical textbook \cite{LL10} gives $\sigma ^{(T)}\sim
\mathrm{const}\,(\Theta /T)^{5}$ (here $\Theta $ is the Debye temperature).
Some other theories predict the dependence $\sim T^{n}$ with $1<n\leq 5$. In
practise, different values of $n$ correspond to different materials.

The method used above for the calculation of the conductivity may be applied
for the clean enough samples of the semimetal if the temperature is much
smaller than $v_{F}\delta p_{z}=v_{F}E\tau $:
\[
T\ll v_{F}E_{z}\tau \sim \sqrt{\frac{v_{F}E_{z}a^{2}}{\sigma _{t}}}\sim
\sqrt{v_{F}E_{z}}\sim 0.001\, \sqrt{\frac{E_{z}}{1\, \mathrm{V}/\mathrm{cm}}}%
\,\mathrm{eV}\sim 10\, \sqrt{\frac{E_{z}}{1\, \mathrm{V}/\mathrm{cm}}}\,
\mathrm{K}
\]%
We also need, that the obtained value of the resistivity $1/\sigma $ is
essentially larger than the resistivity $R^{(0)}$ caused by the interaction
with phonons and impurities to be calculated using the more conventional
methods. In practise, for the temperatures smaller than $\sim 10$ K the
contribution of phonons is already not relevant. As for the impurities,
their contribution to the resistivity may be estimated as
\[
1/\sigma _{i}\sim \frac{3\,(2\pi )^{3}}{2\,|S_{F}|\,l_{i}}
\]%
where $|S_{F}|$ is the area of the Fermi surface while $l_{i}$ is the mean
path corresponding to the scattering on impurities.

For large Fermi surfaces $S_F \sim 4 \pi (\pi/a)^2$ and $\sigma_i \sim \frac{%
l_i}{a^2} \sim \frac{l_i}{a} \, 10^3 $ eV $\sim \frac{l_i}{a} \, 10^4/(%
\mathrm{Ohm} \times \mathrm{cm})$. One can see, that we need rather clean
materials (or rather large electric fields) with
\[
\frac{l_i}{a} \gg \sqrt{\frac{1\, \mathrm{V}/\mathrm{cm}}{E_z}}\times 10^{5}
\]
Notice, that the density of impurities is related to the value of $l_i$ as $%
\rho_i \sim \frac{1}{l_i \sigma^{(i)}_t}$, where $\sigma_t^{(i)}$ is the
transport cross section for the scattering of the quasiparticles on the
impurity. We come to the condition
\[
\Big(\frac{1}{a \sigma_t^{(i)} \rho_i}\Big)^{1/3} \gg \Big(\frac{1\, \mathrm{%
V}/\mathrm{cm}}{E_z}\Big)^{1/6}\times 50
\]
The detailed consideration of this condition depends on the type of the
material and the type of the impurity. Assuming as an example, that the
impurity does not disturb strongly the effective tight - binding model, so
that the cross section $\sigma_t^{(i)}$ corresponds to the effective
distance of the order of the interatomic distance, we come to the
requirement, that there is less than one atom of impurities within each cube
of the original lattice of the size (in lattice units) $100 \times 100
\times 100$.

Under these conditions the considered above resistivity as well as the non -
linear dependence of electric current on the electric field may be observed
experimentally at the temperature of the order of several Kelvin. Recall
also, that the type I Weyl semimetal has the vanishing value of conductivity
at zero temperature.

\section{The type I semimetal in the presence of elastic deformations}

\label{SectGravI}

The low energy of the type I Weyl semimetal is described by the actions Eqs.
(\ref{SHe_3sW2}) and (\ref{SHe_3sW6}). The vierbein $e_{\mu }^{k}$ deviates
from the expression given by Eq. (\ref{Connection00}), while the positions
of the Weyl points $A_{\mu }$ deviate from their unperturbed values. As a
result the fermionic quasiparticles experience the emergent gauge field $%
A_{\mu }(x)$ and the emergent gravity given by $e_{\mu }^{k}(x)$. Both
fields depend on the tensor of elastic deformations. In the presence of
elastic deformations, in principle, the vielbeins (as well as the emergent
gauge fields) may differ for the (left - handed or the right - handed)
fermions incident at the different Fermi points. Let us introduce the tensor
of elastic deformations $u^{ij}=\frac{1}{2}{\LARGE \ }\left( \partial
_{i}u^{j}+\partial _{j}u^{i}\right) $, where $u^{i}$ is the displacement
vector. In the general case the emergent vielbein is expressed up to the
terms linear in displacement as follows\cite{Z2015}: \ {\LARGE \ }
\begin{eqnarray}
e_{a}^{i} &=&\hat{f}_{a}^{i}(1+\frac{1}{3}\gamma _{kij}^{k}u^{ij})-\hat{f}%
_{a}^{n}\gamma _{njk}^{i}u^{jk}  \nonumber \\
e_{0}^{i} &=&-\frac{1}{v_{F}}\gamma _{0jk}^{i}u^{jk},\quad e_{a}^{0}=0
\nonumber \\
e_{0}^{0} &=&\frac{1}{v_{F}}(1+\frac{1}{3}\gamma _{kij}^{k}u^{ij})  \nonumber
\\
|\mathbf{e}| &=&v_{F}(1-\frac{1}{3}\gamma _{kij}^{k}u^{ij})  \nonumber \\
&&a,i,j,k,n=1,2,3  \label{eg}
\end{eqnarray}%
The emergent gauge field is given by
\begin{eqnarray}
A_{i} &\approx &\frac{1}{a}\beta _{ijk}u^{jk},  \nonumber \\
A_{0} &=&0,\quad i,j,k=1,2,3  \label{DiracPosition1_}
\end{eqnarray}%
Here tensors $\beta $ and $\gamma $ may also, in principle, be different for
the right - handed and the left - handed fermions. Following \cite%
{Chernodub:2015wxa} we may assume, that the values of these parameters are
of the order of unity.

Let us discuss the particular tight - binding model of Eq. (\ref{WST}). We
introduce into this model the hopping parameters $t$, $f$, and $r$ and
consider the corresponding tight - binding Hamiltonian:
\begin{equation}
H_{0}=\frac{1}{2}\sum_{\mathbf{x},j=1,2,3}\bar{\psi}(\mathbf{x}+\mathbf{l}%
_{j})\,i\,t\sigma ^{j}\psi (\mathbf{x})-\frac{r}{2}\sum_{\mathbf{x},j=1,2}%
\bar{\psi}(\mathbf{x}+\mathbf{l}_{j})\, \sigma ^{3}\psi (\mathbf{x})+{\,r}%
\sum_{\mathbf{x}}\bar{\psi}(\mathbf{x})\, \sigma ^{3}\psi (\mathbf{x})-\frac{%
i\,f}{2}\sum_{\mathbf{x}}\bar{\psi}(\mathbf{x}+\mathbf{l}_{3})\psi (\mathbf{x%
})+(h.c.)  \label{H-1}
\end{equation}%
Here the sum is over the positions $\mathbf{x}$ of the 3D cubic lattice and
over $j=1,2,3$. $\gamma ^{i}$ are the Dirac matrices in chiral
representation. Vectors $\mathbf{l}_{j}$ connect the nearest neighbor sites
of the lattice. It is assumed that the values of the parameters are chosen
in such a way, that the system remains in the domain of the type I Weyl
semimetal. The conditions to be fulfilled in order to drop into the type II
phase will be considered below in Sect. \ref{SectIIelastic}.

Let us consider the Hamiltonian in momentum representation:
\begin{equation}
H=t\sum_{\mathbf{p},j=1,2,3}\bar{\psi}(\mathbf{p})\, \mathrm{sin}(\mathbf{p}%
\mathbf{l}_{j})\, \sigma ^{j}\psi (\mathbf{p})+r\sum_{\mathbf{p},j=1,2}\bar{%
\psi}(\mathbf{p})\,(1-\mathrm{cos}(\mathbf{p}\mathbf{l}_{j}))\, \sigma
^{3}\psi (\mathbf{p})-f\sum_{\mathbf{p}}\bar{\psi}(\mathbf{p})\, \mathrm{sin}%
(\mathbf{p}\mathbf{l}_{3})\, \psi (\mathbf{p})  \label{H0}
\end{equation}%
Here $\mathbf{p}$ is momentum of the quasiparticle. Next, let us define
\begin{equation}
\hat{p}_{i}=\frac{\mathrm{sin}\,(p_{i}\,a)}{a},\quad k_{i}=p_{i}\,a
\end{equation}%
(where $a=|\mathbf{l}_{i}|$ is the lattice spacing). This gives the one -
particle Hamiltonian
\begin{equation}
H=v_{F}\sum_{k=1,2,3}\sigma ^{k}\hat{p}_{k}+r\sigma ^{3}\Big(\sum_{k=1,2}(1-%
\mathrm{cos}\,(p_{k}a))\Big)-v\, \hat{p}_{3}  \label{WST2}
\end{equation}%
We introduce the dimensionless parameters
\begin{equation}
v_{F}=t\,a,\quad v=f\,a
\end{equation}%
$v_{F}$ has the meaning of Fermi velocity.


\label{SectHop}

In the presence of elastic deformations \cite{ref:LL} the hopping parameters
depend on direction and on the position in space. First let us consider the
simplest model that relates hopping parameters with the tensor of elastic
deformations. In this model the hopping parameter corresponding to the jump
between the two sites $\mathbf{x}$ and $\mathbf{x}+\mathbf{l}_{j}$ depends
only on the real distance between these two sites given by ${r}(\mathbf{x},%
\mathbf{l}_{j})=|\mathbf{l}_{j}+\mathbf{u}(\mathbf{x}+\mathbf{l}_{j})-%
\mathbf{u}(\mathbf{x})|$, where $\mathbf{u}$ is the displacement vector.
Therefore, we substitute the hopping parameter at link $(\mathbf{x},j)$ with
(summation over $k$ and $m$ is assumed): \ \ \ \ \ \ \
\begin{equation}
t\rightarrow t(1-\beta _{t}\,l_{j}^{k}l_{j}^{m}u_{km})=t(1-\beta _{t}u_{jj})
\end{equation}%
and
\begin{equation}
r\rightarrow r(1-\beta _{r}\,l_{j}^{k}l_{j}^{m}u_{km})=r(1-\beta _{r}u_{jj})
\end{equation}%
and
\begin{equation}
f\rightarrow f(1-\beta _{v}\,l_{3}^{k}l_{3}^{m}u_{km})=v(1-\beta _{f}u_{33})
\end{equation}
$\beta _{r}$ and $\beta _{t}$ are the material parameters.

One might also consider the following complication. Let us assume, that the
parameter $t$ receives the extra correction due to the non - diagonal
elements of the deformation tensor \cite{WeylTightBinding}:
\begin{equation}
t\sigma ^{j}\rightarrow t\sigma ^{j}(1-\beta
_{t}\,l_{j}^{k}l_{j}^{m}u_{km})+t\beta _{t}^{\prime }\sum_{n\neq
j}l_{j}^{k}l_{n}^{m}u_{km}\sigma ^{n}=t\sigma ^{j}(1-\beta
_{t}u_{jj})+t\beta _{t}^{\prime }\sum_{n\neq j}u_{jn}\sigma ^{n}
\end{equation}%
with the new material parameter $\beta _{t}^{\prime }$. The modified tight -
binding model corresponds to the Hamiltonian
\begin{eqnarray}
H &=&\frac{1}{2}\sum_{\mathbf{x},j=1,2,3}\bar{\psi}(\mathbf{x}+\mathbf{l}%
_{j})\Big(i\,t(1-\beta _{t}u_{jj})\sigma ^{j}+it\beta _{t}^{\prime
}\sum_{n\neq j}u_{jn}\sigma ^{n}\Big)\psi (\mathbf{x})  \nonumber \\
&&+\frac{1}{2}\sum_{\mathbf{x},j=1,2}\bar{\psi}(\mathbf{x}+\mathbf{l}_{j})%
\Big(-r(1-\beta _{r}u_{jj})\, \sigma ^{3}\Big)\psi (\mathbf{x})  \nonumber \\
&&+\sum_{\mathbf{x}}\bar{\psi}(\mathbf{x})\Big(\,r\,(1-\frac{\beta _{r}}{2}%
u^{(2)})\, \sigma ^{3}\Big)\psi (\mathbf{x})-\frac{1}{2}\sum_{\mathbf{x}}%
\bar{\psi}(\mathbf{x}+\mathbf{l}_{3})\Big(i\,f(1-\beta _{f}u_{33})\Big)\psi (%
\mathbf{x})+(h.c.)
\end{eqnarray}%
We denote
\begin{equation}
u^{(2)}\equiv \sum_{j=1,2}u_{jj},\quad u\equiv \sum_{j=1,2,3}u_{jj}
\end{equation}%
Notice, that the term $r\,(1-\frac{\beta _{r}}{2}u^{(2)})\, \sigma ^{3}$ is
needed in order to save the Fermi points. Without this coherent modification
of the hopping parameter $r$ the fermions become gapped. We arrive at the
one - particle Hamiltonian
\begin{eqnarray}
H &=&v_{F}\sum_{i=1,2,3}\sigma ^{i}\Big(\hat{p}_{i}(1-\beta
_{t}u_{ii})+\beta _{t}^{\prime }\sum_{n\neq i}u_{in}\hat{p}_{n}\Big)
\nonumber \\
&&+r\, \sum_{i=1,2}(1-\beta _{r}u_{ii})\Big[1-\mathrm{cos}\, \mathbf{p}%
\mathbf{l}_{i}\Big]\sigma ^{3}-v(1-\beta _{f}u_{33})\hat{p}_{3}  \label{H2}
\end{eqnarray}%
Here the product of momentum operator $\hat{p}_{k}$ and the coordinate
dependent function $u_{ij}(\mathbf{x})$ is defined as $\frac{1}{2}(\hat{p}%
_{k}u_{ij}(\mathbf{x})+u_{ij}(\mathbf{x})\hat{p}_{k})$. The same symmetric
rule is applied to the product of $\mathrm{cos}\,(p_{k}a)$ and $u_{ij}(%
\mathbf{x})$.

One can see, that for sufficiently small values of $v$, when the system
remains in the type I domain the positions of the Weyl points are not
changed and are given by
\begin{equation}
\mathbf{K}^{+}=\Big(0,0,0\Big),\quad \mathbf{K}^{-}=\Big(0,0,\pi/a \Big)
\end{equation}%
That means, that the emergent gauge field does not appear. In the small
vicinity of $\mathbf{K}_{+}$ the Hamiltonian receives the form
\[
\mathcal{H}^{(R)}={\sigma }^{a}{f}_{a}^{+\,k}(p_{k}-K_{k}^{+}),\ \ \ \
\left( a=0,1,2,3)\right)
\]%
while in the vicinity of $\mathbf{K}^{-}$:
\[
\mathcal{H}^{(L)}=-\sigma ^{3}{f}_{a}^{-\,k}{\sigma }^{a}(p_{k}-K_{k}^{-})%
\sigma ^{3},
\]%
where
\begin{equation}
f_{4\times 3}^{\pm }=v_{F}\left(
\begin{array}{ccc}
0 & 0 & \mp \frac{v}{v_{F}}(1-\beta _{f}u_{33}) \\
1-\beta _{t}u_{11} & \beta _{t}^{\prime }u_{12} & \beta _{t}^{\prime }u_{13}
\\
\beta _{t}^{\prime }u_{21} & 1-\beta _{t}u_{22} & \beta _{t}^{\prime }u_{23}
\\
\beta _{t}^{\prime }u_{31} & \beta _{t}^{\prime }u_{32} & 1-\beta _{t}u_{33}%
\end{array}%
\right)
\end{equation}%
The vierbein $e_{a}^{k}$ is related to tensor $f$ as follows:
\begin{equation}
1=|\mathrm{det}_{4\times 4}e|^{-1}e_{0}^{0},\quad |\mathrm{det}_{4\times
4}e|^{-1}e_{a}^{k}=f_{a}^{k}
\end{equation}%
Here
\[
\mathrm{det}_{4\times 4}e=e_{0}^{0}\, \mathrm{det}_{3\times 3}e
\]%
Therefore, $\mathrm{det}_{3\times 3}e=1$ and
\[
e_{0}^{0}=\mathrm{det}^{-1/3}f_{3\times 3},\quad {e}_{a}^{k}=\frac{f_{a}^{k}%
}{\mathrm{det}^{1/3}f_{3\times 3}},\quad a=0,1,2,3\,,\,k=1,2,3
\]%
Therefore, the $4\times 4$ matrix of the vierbein is
\begin{equation}
e^{\pm }=\left(
\begin{array}{cccc}
\frac{1}{v_{F}}({1+\frac{\beta _{t}}{3}u}) & 0 & 0 & \mp \frac{v}{v_{F}}(1+%
\frac{\beta _{t}}{3}u-\beta _{f}u_{33}) \\
0 & 1+\frac{\beta _{t}}{3}u-\beta _{t}u_{11} & \beta _{t}^{\prime }u_{12} &
\beta _{t}^{\prime }u_{13} \\
0 & \beta _{t}^{\prime }u_{21} & 1+\frac{\beta _{t}}{3}u-\beta _{t}u_{22} &
\beta _{t}^{\prime }u_{23} \\
0 & \beta _{t}^{\prime }u_{31} & \beta _{t}^{\prime }u_{32} & 1+\frac{\beta
_{t}}{3}u-\beta _{t}u_{33}%
\end{array}%
\right)
\end{equation}

\section{The type II Weyl semimetal in the presence of elastic deformations}

\label{SectElasticII}

For the type II Weyl semimetals the low energy theory describes excitations
that reside near the whole Fermi surface, not only near the type II Weyl
points. Therefore we cannot restrict ourselves by the consideration of the
vicinities of the Weyl points, and are to consider the whole momentum space.
In the absence of elastic deformations the type II semimetal appears for $%
v>v_{F}$. In the presence of elastic deformations the corresponding
condition is
\begin{equation}
\left \vert \left(
\begin{array}{ccc}
1-\beta _{t}u_{11} & \beta _{t}^{\prime }u_{12} & \beta _{t}^{\prime }u_{13}
\\
\beta _{t}^{\prime }u_{21} & 1-\beta _{t}u_{22} & \beta _{t}^{\prime }u_{23}
\\
\beta _{t}^{\prime }u_{31} & \beta _{t}^{\prime }u_{32} & 1-\beta _{t}u_{33}%
\end{array}%
\right) ^{-1}\left(
\begin{array}{c}
0 \\
0 \\
\mp \frac{v}{v_{F}}(1-\beta _{f}u_{33})%
\end{array}%
\right) \right \vert >1
\end{equation}%
that is (up to the terms linear in the deformation tensor)
\begin{equation}
v|(1+\beta _{t}u_{33}-\beta _{f}u_{33})|>v_{F}
\end{equation}%
The dispersion of the quasiparticles is given by
\begin{eqnarray}
\mathcal{E} &=&\pm \frac{v_{F}}{a}\Big(\sum_{k=1,2}\Big((1-\beta
_{t}u_{kk})\, \mathrm{sin}\,({p}_{k}a)+\beta _{t}^{\prime }\sum_{n\neq
k}u_{kn}\mathrm{sin}\,({p}_{n}a)\Big)^{2}  \nonumber \\
&&+\Big((1-\beta _{t}u_{33})\, \mathrm{sin}\,({p}_{3}a)+\beta _{t}^{\prime
}\sum_{n\neq 3}u_{3n}\mathrm{sin}\,({p}_{n}a)+\gamma \sum_{k=1,2}(1-\beta
_{r}u_{kk})(1-\mathrm{cos}\,(p_{k}a))\Big)^{2}\Big)^{1/2}  \nonumber \\
&&-\frac{v}{a}\,(1-\beta _{f}u_{33})\, \mathrm{sin}\,({p_{3}}a)  \label{EPS}
\end{eqnarray}%
where
\[
\gamma =\frac{ra}{v_{F}}.
\]

\begin{figure}[h]
\centering
\includegraphics[width=6cm]{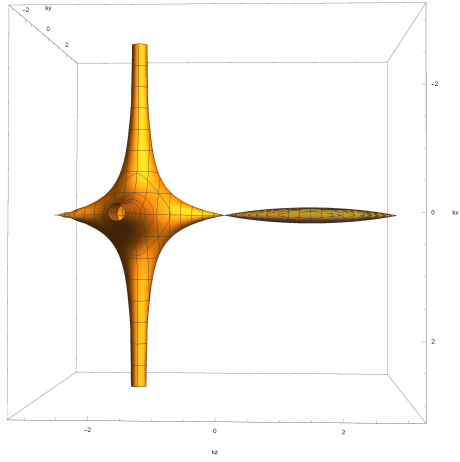}
\caption{The form of the Fermi surface for the type II Weyl semimetal
described by Eq. (\protect\ref{EPS}) (elastic deformations neglected) when
the system is close to the transition to the type I phase. The chosen
parameters are $v_F = 1, v = 1.02$. Here the axes correspond to the values $%
k_x = p_x a, k_y = p_y a, k_z = p_z a$. One can see, that when $v \to v_F +
0 $ the Fermi surface is reduced to the four straight Fermi lines that cross
each other in a single point. When the value of $v$ becomes smaller than $%
v_F $ the Fermi lines disappear and the two Fermi points remain.}
\label{fig4}
\end{figure}

Let us consider how the elastic deformations affect the Fermi surface for
the case
\[
\beta_t = \beta_r = \beta,\quad \beta_t^\prime = \beta_f = 0, \quad \gamma=1
\]
The small variations of the positions of the point on the Fermi surface $%
\delta p_i$, $i=1,2,3$ are related to the small values of the deformation
tensor as follows:
\begin{eqnarray}
0 &=& \Big((1-v^2/v_F^2)\, \mathrm{cos}\,p_3a\, \mathrm{sin}\,p_3a + \mathrm{%
cos}\,p_3\,(2-\mathrm{cos}\,p_1a-\mathrm{cos}\,p_2a) \Big)\, \delta p_3
\nonumber \\
&& + \Big( \mathrm{sin}\,p_1a\,(\mathrm{sin}\,p_3a + 2- \mathrm{cos}\,p_2a) %
\Big)\, \delta p_1  \nonumber \\
&& + \Big( \mathrm{sin}\,p_2a\,(\mathrm{sin}\,p_3a + 2- \mathrm{cos}\,p_1a) %
\Big)\, \delta p_2  \nonumber \\
&& - \frac{\beta}{a} \Big(\mathrm{sin}^2\,p_1a \,u_{11}+ \mathrm{sin}%
^2\,p_2a \,u_{22}  \nonumber \\
&&+ (\mathrm{sin}\,p_3a + 2 - \mathrm{cos}\,p_1a- \mathrm{cos}%
\,p_2a)(u_{33}\,\mathrm{sin}\,p_3a + u_{11}\,(1 - \mathrm{cos}\,p_1a)+
u_{22}\,(1- \mathrm{cos}\,p_2a)) \, \Big)
\end{eqnarray}
Shift of the Fermi surface is given by vector (normal to its original form)
\begin{equation}
\left(%
\begin{array}{c}
\delta p_3 \\
\delta p_1 \\
\delta p_2%
\end{array}
\right) = \left(%
\begin{array}{c}
\Big((1-v^2/v_F^2)\, \mathrm{cos}\,p_3\, \mathrm{sin}\,p_3 + \mathrm{cos}%
\,p_3a\,(2-\mathrm{cos}\,p_1a-\mathrm{cos}\,p_2a) \Big) \\
\Big( \mathrm{sin}\,p_1a\,(\mathrm{sin}\,p_3a + 2- \mathrm{cos}\,p_2a) \Big)
\\
\Big( \mathrm{sin}\,p_2a\,(\mathrm{sin}\,p_3a + 2- \mathrm{cos}\,p_1a) \Big)%
\end{array}
\right)\delta \xi  \label{CFFS}
\end{equation}
where
\begin{eqnarray}
\delta \xi[u_{11},u_{22},u_{33}] &=&\frac{\beta}{a} \Big[\Big(%
(1-v^2/v_F^2)\, \mathrm{cos}\,p_3a\, \mathrm{sin}\,p_3 a+ \mathrm{cos}%
\,p_3a\,(2-\mathrm{cos}\,p_1a-\mathrm{cos}\,p_2a) \Big)^2  \nonumber \\
&& + \Big( \mathrm{sin}\,p_1a\,(\mathrm{sin}\,p_3a + 2- \mathrm{cos}\,p_2a) %
\Big)^2  \nonumber \\
&& + \Big( \mathrm{sin}\,p_2a\,(\mathrm{sin}\,p_3a + 2- \mathrm{cos}\,p_1a) %
\Big)^2\Big]^{-1}\times  \nonumber \\
&& \Big[\mathrm{sin}^2\,p_1a \,u_{11}+ \mathrm{sin}^2\,p_2a \,u_{22}
\nonumber \\
&&+ (\mathrm{sin}\,p_3a + 2 - \mathrm{cos}\,p_1a- \mathrm{cos}%
\,p_2a)(u_{33}\,\mathrm{sin}\,p_3a + u_{11}\,(1 - \mathrm{cos}\,p_1a)+
u_{22}\,(1- \mathrm{cos}\,p_2a)) \, \Big]  \label{CFFS2}
\end{eqnarray}

If the free energy that is responsible for elastic deformations is
isotropic, it is given by
\begin{equation}
F=\frac{1}{2}\int d^{3}x\Big(\lambda u^{2}+2\mu u_{ik}u_{ik}\Big)
\end{equation}%
where $\lambda $ and $\mu $ are the Lame coefficients. Minimum of this
functional gives the elasticity equations for the static deformations. In
order to describe the deformations depending on time we should consider the
action of the form
\begin{equation}
S=\frac{1}{2}\int d^{3}xdt\Big(\rho (\partial _{t}{u_{i}})^{2}-\lambda
u^{2}-2\mu u_{ik}u_{ik}\Big)  \label{Selastic}
\end{equation}%
where $\rho $ is density.

In order to build the effective theory for the deviation of the Fermi
surface from its original form we describe this deviation by the above
introduced parameter $\delta \xi$. It becomes the effective scalar field $z$
(depending on time) defined on the Fermi surface parametrized by the two
coordinates $s_k$, $k = 1,2$. The corresponding partition function is
\begin{equation}
Z = \int D z(s_1,s_2,t) \delta \Big(z - \delta \xi[u] \Big) Du D\bar{\psi}
D\psi e^{iS[u] + iS_f[\bar{\psi},\psi,u]}
\end{equation}
where $S_f$ is the fermion action. The effective action is
\begin{equation}
e^{i S_0[z]}\equiv \int \delta \Big(z - \delta \xi[u] \Big) Du D\bar{\psi}
D\psi e^{iS[u] + iS_f[\bar{\psi},\psi,u]}
\end{equation}
We may assume here, that the integral over the fermions in the leading order
contributes the effective action through the modification of the parameters
of Eq. (\ref{Selastic}). Then we have
\begin{equation}
e^{i S_0[z]} = \int \delta \Big(z - \delta \xi[u] \Big) Du e^{iS[u]}
\end{equation}

So far we discussed how the elastic deformation alters the position of the
Fermi surface. However, we did not discuss the change in the occupation of
the states inside the modified Fermi surface. There are the two distinct
limiting regimes. In the first case there is enough time for the fermions to
rearrange themselves and occupy the states with the negative energy bounded
by the modified Fermi surface. This occurs if the relaxation to thermal
equilibrium occurs much faster than the variation of the Fermi surface
shape. In this case if the elastic deformation has the form of the
propagating wave we may speak of the special mode of the zero sound, which
is the propagating vibration of the Fermi surface. Unlike the conventional
zero sound it is semi - equilibrium phenomenon from the point of view of the
ensemble of the fermions. Recall that the conventional zero sound occurs due
to the oscillation of the shape of the region of the occupied states in
momentum space around equilibrium. It is, therefore, the completely non -
equilibrium phenomenon that is described by kinetic theory (see, for
example, chapter 4 of \cite{LL9}). The conventional zero sound velocity
cannot exceed the Fermi velocity (otherwise, the wave disappears fast due to
the dissipation). However, there is no such bound on the special zero sound
modes discussed here.

The condition for the realization of this regime is that the modification of
the Fermi surface occurs more slowly than the relaxation of the ensemble of
the fermionic quasiparticles to their equilibrium. That means that $\tau
\frac{d}{dt} \delta p$ is much smaller than ${\delta p}$ (here $\delta p$ is
the shift of the Fermi surface, $\tau$ is the typical inter - collision
time). The latter may be estimated through the mean free path
\[
l\sim v_{F}\tau \approx \frac{1}{\rho \sigma _{t}}
\]
where $\rho$ is the density of the quasiparticles outside of the Fermi
surface. It may be estimated as
\[
\rho \sim \frac{\frac{d}{dt} \delta p}{4\pi }\int \frac{dp_{x}dp_{y}}{(2\pi
)^{2}} \tau \sim \frac{\frac{d}{dt} \delta p}{4\pi a^2 } \tau
\]
which gives
\[
\tau \sim \frac{1}{\sqrt{v_F \frac{d}{dt} \delta p}}
\]
In turn, for the rate of the modification of the Fermi surface caused by the
sound wave $u \sim U \,\mathrm{cos}(\omega t - \mathbf{q} \mathbf{x} )$ we
have
\[
\delta p \sim \frac{1}{a} U \,|\mathbf{q}|, \quad \frac{d}{dt} \delta p \sim
\frac{\omega}{a} U \,|\mathbf{q}|
\]
Here $\omega \sim v_{sound} |\mathbf{q}|$ where $v_{sound}$ is the speed of
sound while $\mathbf{q}$ is the wave vector. The sound wavelength is $%
\lambda_{sound} \sim \frac{1}{|\mathbf{q}|}$. We come to the condition: $%
\sqrt{\frac{\omega}{v_F a} U \,|\mathbf{q}| } \ll \frac{1}{a} U \,|\mathbf{q}%
| $ that is
\begin{equation}
\frac{v_{sound}}{v_F}a \ll U  \label{soundcond0}
\end{equation}
Here $U$ is the typical displacement of atoms due to the elastic
deformations in the sound wave. The speed of sound is typically of the order
of $3\times 10^3$ m$/$s. At the same time the typical value of the Fermi
velocity in Weyl semimetal is rather high - of the order of $10^{-2} c \sim
3\times 10^6$ m$/$s. The elasticity theory requires $U \ll \lambda_{sound}$,
and we need
\begin{equation}
10^{-3} a \ll U \ll \lambda_{sound}  \label{soundcond}
\end{equation}

In the opposite limiting case strictly speaking we already cannot speak of
the zero sound because the modification of the (would be) Fermi surface
occurs much faster than the relaxation to the thermal equilibrium.
Obviously, this occurs only for the sound waves with very small amplitude $%
\frac{v_{sound}}{v_F}a \gg U$. For such vibrations of atoms almost all
originally occupied states remain occupied. However, the wave of the
vibration of atoms is able to excite the fermionic quasiparticles due to the
interaction between them.

Quantization of the sound waves gives the typical value of the amplitude in
the wave corresponding to one phonon that is much smaller than the above
discussed bound $\frac{v_{sound}}{v_F}a$. Therefore, on the quantum level
the single phonons do not produce the zero sound wave of the type discussed
here. Those waves correspond to the classical sound wave, which contains the
huge number of single phonons.

\section{Relation between the sound waves of elasticity and the special
modes of the zero sound}

\label{SectZeroSoundPhonons}

The action Eq. (\ref{Selastic}) describes the two types of the sound waves -
the longitudinal ones with velocity $v_\parallel = \sqrt{\frac{2 \mu +
\lambda}{\rho}}$ and the transversal ones with velocity $v_\bot = \sqrt{%
\frac{\mu }{\rho}}$. On the quantum level those waves become the two types
of phonon excitations with the linear dispersion and the same values of
velocity. If the condition of Eq. (\ref{soundcond}) is satisfied and the
wavelength is sufficiently large, then each sound wave produces the
vibration of the Fermi surface that propagates with the same velocity. It is
described by the variation of the form of the Fermi surface given by Eqs. (%
\ref{CFFS}), (\ref{CFFS2}), where we should substitute the deformation
tensor $u_{ij}$ corresponding to the three modes of the sound waves. Those
modes appear from the Fourier transformation of $u$:
\begin{equation}
\mathbf{u} (t,\mathbf{x}) = \sqrt{V/T} \int \frac{d^3 \mathbf{q} d E}{%
(2\pi)^4} \Big( U^{\mathbf{q},E}_\parallel \frac{\mathbf{q}}{|\mathbf{q}|}\,
\mathrm{exp}\Big(-i E t + i \mathbf{q}\, \mathbf{x}\Big) + \sum_{a=1,2} U^{%
\mathbf{q},E}_{\bot,a} \hat{\mathbf{m}}^{\mathbf{q}}_a \,\mathrm{exp}\Big(-i
E t + i \mathbf{q}\, \mathbf{x}\Big) \Big)
\end{equation}
Here $\hat{\mathbf{m}}^{\mathbf{q}}_a$ for $a=1,2$ are the two unity vectors
orthogonal to $\mathbf{q}$. We assume, that $\hat{\mathbf{m}}^{\mathbf{q}}_a
= - \hat{\mathbf{m}}^{-\mathbf{q}}_a$ and find that $U^{\mathbf{q}%
,E}_\parallel = - [U^{-\mathbf{q},-E}_\parallel ]^+$, which guarantees that $%
\mathbf{u} (t,\mathbf{x})$ is real. $V$ is the overall volume while $T$ is
temperature. The effective action for the variations of the Fermi surface
(the zero sound) receives the form
\begin{eqnarray}
e^{i S_0[z]} &=& \int \delta(z - \delta \xi(u)) \Pi_{\mathbf{q}}\Pi_E d U^{%
\mathbf{q},E}_\bot d U^{\mathbf{q},E}_\parallel \mathrm{exp}\Big[ \frac{i V}{%
2T}\int \frac{d^{3}\mathbf{q}d E}{(2\pi)^4} \rho \sum_{a=1,2} [ U^{\mathbf{q}%
}_{\bot,a}]^+(E^2 - \omega^2_\bot(\mathbf{q}))U^{\mathbf{q}}_{\bot,a}
\nonumber \\
&& + \frac{i V}{2T}\int \frac{d^{3}\mathbf{q}d E}{(2\pi)^4} \rho [ U^{%
\mathbf{q}}_\parallel]^+(E^2 - \omega^2_\parallel(\mathbf{q}))U^{\mathbf{q}%
}_\parallel \Big]
\end{eqnarray}
where $\omega_\bot(\mathbf{q}) = v_\bot |\mathbf{q}|$ and $\omega_\parallel(%
\mathbf{q}) = v_\parallel |\mathbf{q}|$.

One may represent
\begin{equation}
\delta \zeta (p_1,p_2,p_3;t,\mathbf{x}) = \mathcal{K}^{ij}(p_1,p_2,p_3)
\partial_j {u}_i(t,\mathbf{x}) = \sqrt{V/T} \int \frac{d^3 \mathbf{q} d E}{%
(2\pi)^4} \delta \zeta^{\mathbf{q},E}(p_1,p_2,p_3) \, \mathrm{exp}\Big(-i E
t + i \mathbf{q}\, \mathbf{x}\Big)
\end{equation}
where $\mathcal{K}^{ij}(p_1,p_2,p_3) \partial_j $ is the linear differential
operator defined by Eq. (\ref{CFFS2}). For the Fourier modes $\delta \zeta^{%
\mathbf{q},E}= \delta [\zeta^{-\mathbf{q},-E}]^+$ we have
\[
\delta \zeta^{\mathbf{q},E}(p_1,p_2,p_3) =i \mathcal{K}^{ij}(p_1,p_2,p_3)
q_j \Big( U^{\mathbf{q},E}_\parallel \frac{q_i}{|\mathbf{q}|}\, +
\sum_{a=1,2} U^{\mathbf{q},E}_{\bot,a} \hat{ m}^{\mathbf{q}}_{i,a} \, \Big)
\]
where
\begin{eqnarray}
\mathcal{K}^{ij}(p_1,p_2,p_3) &=&\frac{\beta}{a} \Big[\Big((1-v^2/v_F^2)\,
\mathrm{cos}\,p_3a\, \mathrm{sin}\,p_3 a+ \mathrm{cos}\,p_3a\,(2-\mathrm{cos}%
\,p_1a-\mathrm{cos}\,p_2a) \Big)^2  \nonumber \\
&& + \Big( \mathrm{sin}\,p_1a\,(\mathrm{sin}\,p_3a + 2- \mathrm{cos}\,p_2a) %
\Big)^2  \nonumber \\
&& + \Big( \mathrm{sin}\,p_2a\,(\mathrm{sin}\,p_3a + 2- \mathrm{cos}\,p_1a) %
\Big)^2\Big]^{-1}\times  \nonumber \\
&& \Big[\mathrm{sin}^2\,p_1a \,\delta_{i1}\delta_{j1}+ \mathrm{sin}^2\,p_2a
\,\delta_{i2}\delta_{j2}  \nonumber \\
&&+ (\mathrm{sin}\,p_3a + 2 - \mathrm{cos}\,p_1a- \mathrm{cos}%
\,p_2a)(\delta_{i3}\delta_{j3} \,\mathrm{sin}\,p_3a + \delta_{i1}\delta_{j1}
\,(1 - \mathrm{cos}\,p_1a)+ \delta_{i2}\delta_{j2} \,(1- \mathrm{cos}%
\,p_2a)) \, \Big]  \label{CFFS8}
\end{eqnarray}
We also introduce
\begin{equation}
z(p_1,p_2,p_3;t,\mathbf{x}) = \sqrt{V/T} \int \frac{d^3 \mathbf{q} d E}{%
(2\pi)^4} z^{\mathbf{q},E}(p_1,p_2,p_3) \, \mathrm{exp}\Big(-i E t + i
\mathbf{q}\, \mathbf{x}\Big)
\end{equation}
For the Fourier components of $z$ we have the following effective action
\begin{eqnarray}
e^{i S_0[z]} &=& \int \Pi_{p_1,p_2,p_3} \Pi_{\mathbf{q}}\Pi_E \delta\Big(z^{%
\mathbf{q},E}(p_1,p_2,p_3) - i \mathcal{K}^{ij}(p_1,p_2,p_3) q_j \Big( U^{%
\mathbf{q},E}_\parallel \frac{q_i}{|\mathbf{q}|}\, + \sum_{a=1,2} U^{\mathbf{%
q},E}_{\bot,a} \hat{ m}^{\mathbf{q}}_{i,a} \, \Big)\Big)  \nonumber \\
&& d U^{\mathbf{q},E}_\bot d U^{\mathbf{q},E}_\parallel \mathrm{exp}\Big[
\frac{i V}{2T}\int \frac{d^{3}\mathbf{q}d E}{(2\pi)^4} \rho \sum_{a=1,2} [
U^{\mathbf{q}}_{\bot,a}]^+(E^2 - \omega^2_\bot(\mathbf{q}))U^{\mathbf{q}%
}_{\bot,a}  \nonumber \\
&& + \frac{i V}{2T}\int \frac{d^{3}\mathbf{q}d E}{(2\pi)^4} \rho [ U^{%
\mathbf{q}}_\parallel]^+(E^2 - \omega^2_\parallel(\mathbf{q}))U^{\mathbf{q}%
}_\parallel \Big]
\end{eqnarray}
One can see, that each sound wave given by the wave vector $\mathbf{q}$ and
the complex numbers $U^{\mathbf{q}}_\parallel, U^{\mathbf{q}}_{\bot,1}, U^{%
\mathbf{q}}_{\bot,2}$ corresponds to the particular configuration of the
Fourier transform of $z(p_1,p_2,p_3;t,\mathbf{x})$ with respect to $t,
\mathbf{x}$. This is the propagating mode of the zero sound. Its components
proportional to $U^{\mathbf{q}}_\parallel, U^{\mathbf{q}}_{\bot,a}$
propagate correspondingly with the velocities of sound $v_\parallel$ and $%
v_\bot$.

For each wave vector $\mathbf{q}$ there are the three modes of sound (one
longitudinal mode and two transverse modes):
\begin{eqnarray}
\mathbf{u}^{\mathbf{q}}_\parallel (t,\mathbf{x}) & = & \mathrm{Re} \, U^{%
\mathbf{q},E}_\parallel \frac{\mathbf{q}}{|\mathbf{q}|}\, \mathrm{exp}\Big(%
-i \omega_{\parallel}(\mathbf{q}) t + i \mathbf{q}\, \mathbf{x}\Big)
\nonumber \\
\mathbf{u}^{\mathbf{q}}_{\bot,a} (t,\mathbf{x}) & = & \mathrm{Re} \, U^{%
\mathbf{q},E}_{\bot,a} \hat{\mathbf{m}}^{\mathbf{q}}_a \,\mathrm{exp}\Big(-i
\omega_{\bot}(\mathbf{q}) t + i \mathbf{q}\, \mathbf{x}\Big)  \label{modes}
\end{eqnarray}
Each mode produces the vibration of the Fermi surface with the particular
shape given by Eqs. (\ref{CFFS}), (\ref{CFFS2}) (we should substitute into
those equations the corresponding expression from Eq. (\ref{modes})). If the
temperature is sufficiently small
\[
T\ll v_{F}E_{z}\tau \sim 10\, \sqrt{\frac{E_{z}}{1\, \mathrm{V}/\mathrm{cm}}}%
\, \mathrm{K}
\]
then the fermionic quasiparticles feel the discussed here zero sound waves
as the waves of the deformation of the Fermi surface.

\section{The type II Weyl semimetal in the presence of elastic deformations
and the conductivity}

\begin{figure}[h]
\centering
\includegraphics[width=6cm]{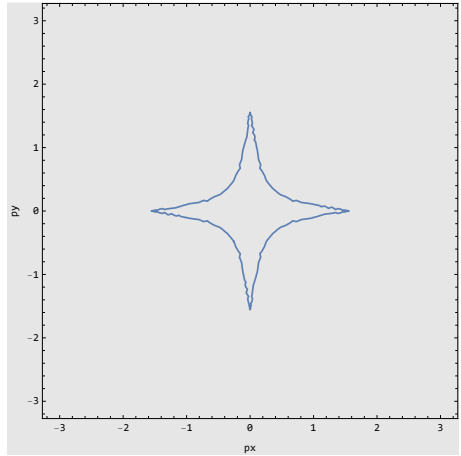}
\caption{The typical form of the slice of the Fermi surface at $p_z a = -%
\protect\pi/2$ for $v_F = v$, small but positive $u_{33}$ with the remaining
components of the deformation tensor that are equal to zero.}
\label{fig12}
\end{figure}

\label{SectIIelastic}

The above expressions allow to estimate the contribution to the electric
current of Eq. (\ref{jE2}) due to the elastic deformations:
\begin{equation}
j_{z}=\sqrt{\frac{2E_{z}}{v_{F}\sigma _{t}}}\frac{\int \frac{dp_{x}dp_{y}}{%
(2\pi )^{3}}\Big|\frac{\partial \mathcal{E}}{\partial p_{z}}\Big|}{\Big[\int
\frac{dp_{x}dp_{y}}{(2\pi )^{3}}\Big]^{1/2}}  \label{jE24}
\end{equation}
where we should substitute the expression of Eq. (\ref{EPS}) for $\mathcal{E}
$ and evaluate it in a vicinity of the Fermi surface:
\begin{eqnarray}
\frac{\partial \mathcal{E}}{\partial p_{z}}&=& \frac{v^2_{F}}{v a}(1-\beta
_{t}u_{33})\, \mathrm{\cos }\,{p}_{3}a\frac{\left[ (1-\beta _{t}u_{33})\,
\mathrm{sin}\,{p}_{3}a+\beta _{t}^{\prime }\sum_{n\neq 3}u_{3n}\mathrm{sin}\,%
{p}_{n}a+\gamma \sum_{k=1,2}(1-\beta _{r}u_{kk})(1-\mathrm{cos}\,p_{k}a)%
\right] }{(1-\beta _{f}u_{33})\, \mathrm{\sin }{p_{3}a}}  \nonumber \\
&&-\frac{v}{a}\,(1-\beta _{f}u_{33})\, \mathrm{\cos }{p_{3}a}  \nonumber \\
&\approx& \frac{v_{F}}{a} \, \frac{v_F}{v}\, \mathrm{\cos }\,({p}_{3}a) %
\Big[ 1-\frac{v^2}{v_F^2}-\Big(2\beta _{t}-\beta_f(1+\frac{v^2}{v_F^2})\Big)%
u_{33} +\beta _{t}^{\prime }\sum_{n\neq 3}u_{3n}\frac{\mathrm{sin}\,({p}%
_{n}a)}{\mathrm{sin} \, (p_3 a)}  \nonumber \\
&&+\gamma \sum_{k=1,2}(1-\beta _{r}u_{kk}+\beta_f u_{33})\frac{(1-\mathrm{cos%
}\,(p_{k}a))}{\mathrm{sin}\,(p_3 a)}\Big]  \label{Eu}
\end{eqnarray}
In addition we should take the integrals in Eq. (\ref{jE24}) over the Fermi
surface of the form changed due to the elastic deformations. The change of
the Fermi surface form is given by Eqs. (\ref{CFFS}), (\ref{CFFS2}). The
expressions for the electric current contain the terms proportional to the
deformation tensor with some coefficients. In general case the expressions
for those coefficients are rather complicated and we do not represent them
here.

Let us notice the case, when our expressions are simplified. This is the
degenerate case, when the system without the elastic deformation is at the
position of the transition between the type II and the type I Weyl
semimetals (i.e. $v = v_F$).

In this intermediate state instead of the Fermi surface there are the Fermi
lines, and the value of electric current at zero temperature vanishes.
However, the elastic deformations may drive further the system towards the
type II phase, and then the electric current at zero temperature appears
given by Eq. (\ref{jE24}), which is simplified considerably. Namely, in the
leading order there is no at all the contribution from the terms containing $%
u_{ij}$ in Eq. (\ref{Eu}). The only contribution remains from the change of
the form of the Fermi surface.

First of all, in the intermediate state in the absence of elastic
deformations we have the three Fermi lines that intersect each other at the
point $(0,0, - \pi/2)$:
\begin{equation}
(III): p_1 = p_2 = 0; \quad (II): p_3 = -\pi/2, p_1=0; \quad (I): p_3 =
-\pi/2, p_2 = 0
\end{equation}
Correspondingly, the velocity of the quasiparticles that reside along the
Fermi lines is given by:
\begin{eqnarray}
&(III)&: \frac{\partial \mathcal{E}}{\partial p_{z}}= \frac{v_{F}}{a} \,
\frac{v_F}{v}\, \mathrm{\cos }\,({p}_{3}a) \Big[ 1-\frac{v^2}{v_F^2}-\Big(%
2\beta _{t}-\beta_f(1+\frac{v^2}{v_F^2})\Big)u_{33} \Big]  \nonumber \\
&&= - 2\frac{v_{F}}{a} \, \mathrm{\cos }\,({p}_{3}a) \Big(\beta _{t}-\beta_f%
\Big)u_{33} \\
&(II)&: \frac{\partial \mathcal{E}}{\partial p_{z}}= 0  \nonumber \\
&(I)&: \frac{\partial \mathcal{E}}{\partial p_{z}}= 0  \nonumber
\end{eqnarray}
Therefore, we will be interested in the modification of the form of the
Fermi surface part reduced to the line $(III)$ without elastic deformations.

Again, we restrict ourselves by the case
\[
\beta _{t}=\beta _{r}=\beta ,\quad \beta _{t}^{\prime }=\beta _{f}=0,\quad
\gamma =1
\]%
In the presence of elastic deformations, outside of the vicinity of the
point $p_{3}=-\pi /(2a),$ the Fermi surface has the form of a small tube
surrounding the $p_{3}$ axis. It is given by equation (see the discussion in
the preceding Section)
\begin{eqnarray}
0 &=&\Big((1-\beta u_{11})^{2}+\mathrm{sin}(p_{3}a)(1-\beta u_{11})(1-\beta
u_{33})\Big)p_{1}^{2}  \nonumber \\
&&+\Big((1-\beta u_{22})^{2}+\mathrm{sin}(p_{3}a)(1-\beta u_{22})(1-\beta
u_{33})\Big)p_{2}^{2}-2\mathrm{sin}^{2}(p_{3}a)\beta u_{33}/a^{2}.
\end{eqnarray}%
We keep the terms linear in elastic deformations and obtain
\begin{eqnarray}
2\,\mathrm{sin}^{2}(p_{3}a)\beta u_{33}/a^{2} &=&\Big(1+\mathrm{sin}%
(p_{3}a)-(2+\mathrm{sin}(p_{3}a))\beta u_{11}-\mathrm{sin}(p_{3}a)\beta
u_{33}\Big)p_{1}^{2}  \nonumber \\
&&+\Big(1+\mathrm{sin}(p_{3}a)-(2+\mathrm{sin}(p_{3}a))\beta u_{22}-\mathrm{%
sin}(p_{3}a)\beta u_{33}\Big)p_{2}^{2}
\end{eqnarray}%
One can see, that for $u_{33}>0$ the Fermi surface appears while for $%
u_{33}<0$ the system drops into the type I domain. In the former case each
slice of the Fermi surface, at a particular value of $p_{3}$, has the form
of a circle (for values of $p_{3}$ outside the vicinity of $-\pi /(2a)$).
Moreover, in the leading order in elastic deformations we may consider it as
a circle with radius $\sim \sqrt{u_{33}}$. In the following we restrict
ourselves to the case when the only nonvanishing component of the
deformation tensor is $u_{33}$.

Near the point $p_{3}=-\pi /(2a)$ the form of the Fermi surface at nonzero $%
u_{33}$ is changed. It surrounds the cross connecting the points $%
(p_{1},p_{2},p_{3})=(\pm \pi /(2a),0,-\pi /(2a))$ and $(0,\pm \pi /(2a),-\pi
/(2a))$. When $u_{33}$ is small, the area of this surface is small too. In
Fig. \ref{fig12} we represent the typical form of the slice of the Fermi
surface at $p_z a = -\pi/2$ for $v_F = v$, small but positive $u_{33}$ with
the remaining components of the deformation tensor that are equal to zero.

Both integrals in Eq. (\ref{jE24}) are non - analytical functions of $\beta
u_{33} $ at $\beta u_{33}=0$. Those functions are rather complicated and we
do not represent them here. Instead we make a rough estimate of the order of
magnitude of the electric current:
\begin{eqnarray}
j_{z} &=&\sqrt{\frac{2E_{z}}{v_{F}\sigma _{t}}}\frac{\int \frac{dp_{x}dp_{y}%
}{(2\pi )^{3}}\Big|\frac{\partial \mathcal{E}}{\partial p_{z}}\Big|}{\Big[%
\int \frac{dp_{x}dp_{y}}{(2\pi )^{3}}\Big]^{1/2}}\approx \sqrt{\frac{2E_{z}}{%
(2\pi )^{3}v_{F}\sigma _{t}}}\langle \Big|\frac{\partial \mathcal{E}}{%
\partial p_{z}}\Big|\rangle \Big[\int dp_{x}dp_{y}\Big]^{1/2}  \nonumber \\
&\approx &\frac{1}{a}\sqrt{\frac{8v_{F}E_{z}}{(2\pi )^{3}\sigma _{t}}}\beta
u_{33}\,\langle \Big|\mathrm{cos}(p_{3}a)\Big|\rangle \Big[\int dp_{x}dp_{y}%
\Big]^{1/2}\sim \frac{1}{a}\sqrt{\frac{8v_{F}E_{z}}{(2\pi )^{3}\sigma _{t}}}%
\beta u_{33}\,\frac{2}{\pi }\,(8\sqrt{\beta u_{33}})^{1/2}\,\mathrm{log}%
^{1/2}\,\frac{1}{\beta u_{33}}  \label{jE242}
\end{eqnarray}%
Here we used the estimate
\[
\langle \Big|\mathrm{cos}(p_{3}a)\Big|\rangle \approx \frac{2}{\pi }%
\int_{0}^{\pi /2}\mathrm{cos}(x)dx=\frac{2}{\pi }
\]%
while the integral $\int dp_{x}dp_{y}$ may be fitted by the following
asymptotic expression
\begin{equation}
\int dp_{x}dp_{y}\approx \mathrm{const}\,\times 8(\beta u_{33})^{1/2}\,%
\mathrm{log}\,\frac{1}{\beta u_{33}}  \label{Aest}
\end{equation}%
where $\mathrm{const}$ is of the order of unity. The final order of
magnitude estimate has the form
\begin{equation}
j_{z}\sim \frac{4}{a\pi ^{2}}\sqrt{\frac{2v_{F}E_{z}}{\pi \sigma _{t}}}%
\theta (u_{33})\,(\beta u_{33})^{5/4}\,\mathrm{log}^{1/2}\,\frac{1}{\beta
u_{33}}  \label{jE246}
\end{equation}

This expression demonstrates, that the system with the Fermi lines, where
the DC conductivity vanishes at zero temperature may acquire the huge
conductivity in the presence of elastic deformations because it is driven by
them to the type II state. Namely, we may estimate this conductivity ($\beta
u_{33}\ll 1 $) as follows (compare with Eqs. (\ref{s1}) and (\ref{s2})):
\[
\sigma \sim \sqrt{\frac{1\,V/cm}{E_{z}}}\times 10^{9}\times \theta(u_{33})
(\beta u_{33})^{5/4}\, \mathrm{log}^{1/2}\,\frac{1}{\beta u_{33}} \times
1/(Ohm\times cm)
\]

\section{Conclusions}

\label{SectConcl}

In the present paper we systematically investigate the type II Weyl
semimetals at small temperatures assuming that they do not drop into the
superconducting state. Several phenomena specific for those materials are
considered here based on the particular tight - binding model. Although this
model is too simple to describe quantitatively the real existing materials
we expect that qualitatively the physics of the considered phenomena may be
understood properly using this model.

\begin{enumerate}
\item {} First of all we discuss the chiral anomaly in the type II Weyl
semimetals that may exist without any external magnetic field. In the type I
Weyl semimetals the chiral anomaly appears only in the presence of external
magnetic field. In the presence of magnetic field there is only one lowest
Landau level and as a result (in the case presented on Fig. \ref{fig1})
close to the left - handed Weyl point the left - moving states remain, while
close to the right - handed Weyl point there are only the right - moving
states. The external electric field applied along the line connecting the
Weyl points gives rise to the drift of the occupied states inside momentum
space. As a result the pairs appear to consist of the right - handed
particle and the left - handed hole. This is the ordinary chiral anomaly.

In the type II Weyl materials the Weyl points also always come in pairs. On
Fig. \ref{fig.1} we represent the typical pattern of the type II Weyl
semimetal. Close to one of the two type II-Weyl points all states are right
- moving, while close to the other one all states are left - moving. As a
result, in the presence of the external electric field directed along the
line connecting the Weyl points the states flow in momentum space and as a
result pairs appear consisting of the particles that reside in the vicinity
of one of the type II-Weyl points and the holes from the vicinity of the
other one.

This phenomenon may be considered as the chiral anomaly without any external
magnetic field. It may be observed experimentally through the measurement of
conductivity. We find that at sufficiently small temperatures and for
sufficiently clean materials  Ohm's law is broken and the electric current
that appears due to the mentioned above type of the chiral anomaly, is
proportional to the square root of electric field. As a result the
resistivity depends on the electric field as $\sim \sqrt{E_{z}}$. Its
absolute value may exceed the resistivity caused by impurities only for very
clean materials and at very small temperatures.

It is worth mentioning that qualitatively the same behavior of conductivity
and the breakdown of Ohm's law at very small temperatures takes place for
clean ordinary metals (see the estimates in Sect. \ref{SectCondII}).

\item {} Next, we consider the effect of the elastic deformations on the
considered tight - binding model. The specific feature of this model is that
the emergent gauge field does not appear when the model is in the type I
domain. However, the emergent gravitational field appears, and was
calculated.

In the type II domain in our model the elastic deformations affect the form
of the Fermi surface. This deformation of the Fermi surface may propagate
thus giving rise to the special modes of the zero sound. For each wave -
vector there are three sound waves of the elasticity theory. Each of them
gives rise to a propagating variation of the shape of the Fermi surface.
This establishes a one - to - one correspondence between the discussed modes
of the zero sound and the modes of the classical sound waves of the
elasticity theory.

Note that the zero sound modes discussed here are different from the
conventional zero sound waves. The latter are completely non - equilibrium
phenomena and are propagating oscillations of the regions of the occupied
fermionic states (in momentum space). Those oscillations occur without any
relation to the oscillations of the atoms of the crystal lattice. They are
described often by the collisionless Boltzmann kinetic equation (see \cite%
{LL9}).

The zero sound modes considered here are the equilibrium phenomena from the
point of view of the fermionic ensemble. They are forced by the vibrations
of the crystal lattice existing in the form of the sound waves.
Correspondingly, they have the same velocity. We consider them using the
path integral formalism. Although the sound waves giving rise to the zero
sound modes are classical (i.e. contain a huge number of phonons), the
dynamics of the fermions is quantum, and therefore our method is more
appropriate for the description of their dynamics rather than the formalism
of the Boltzmann equation, which by definition, treats the fermionic
quasiparticles semiclassically. In spite of the essential difference between
the conventional zero sound modes and the zero sound modes discussed in the
present paper, it is appropriate to proceed calling them \textit{zero sound,}
thus accepting the definition given in chapter 8.1.5 of \cite{Volovik2003}
rather than the definition of \cite{LL9} and \cite{LL10}. The Fermi surface
itself is not necessarily the notion that occurs in the noninteracting
fermionic system, but rather the surface in momentum space, where the two
point fermionic Green function is singular. This surface is able to
oscillate, and such oscillations are able to propagate. In \cite{Volovik2003}
the zero sound is considered as \textit{any} propagating oscillation of this
surface without any relation to its origin.

Elastic deformations cause a contribution to electric conductivity. We see
that in the general case such a contribution is proportional to the
deformation tensor. The especially interesting case is when the system
without strain is in the intermediate state between the type II and the type
I phases. In this state there are Fermi lines instead of Fermi points or
Fermi surface, and the DC conductivity vanishes at zero temperature.
However, even small elastic deformations may drive the system to the type II
state. Thus a huge conductivity appears proportional to a non - analytical
function of the appropriate component of the deformation tensor.
\end{enumerate}

To conclude, we discussed the properties specific for the type II Weyl
semimetals that manifest themselves at small temperatures. Some of these
properties may be observed experimentally through the measurement of
electric conductivity.

M.A.Z. kindly acknowledges useful discussions with G.E.Volovik and
M.N.Chernodub.


\begin{thebibliography}{99}
\bibitem{semimetal_discovery} Z. K. Liu et al., \textsl{``Discovery of a
Three-dimensional Topological Dirac Semimetal, Na${}_3$Bi''}, Science (2014)
\textbf{343}, 864 [arXiv:1310.0391].

\bibitem{semimetal_discovery2} M. Neupane et al., {``Observation of a
topological 3D Dirac semimetal phase in high-mobility Cd${}_3$As${}_2$''}
Nature Commun. \textbf{05}, 3786 (2014) [arXiv:1309.7892].

\bibitem{semimetal_discovery3} S. Borisenko et al., {``Experimental
Realization of a Three-Dimensional Dirac Semimetal''}, Phys. Rev. Lett.
\textbf{113}, 027603 (2014) [arXiv:1309.7978].

\bibitem{ref:semimetal:3} Z. K. Liu et al., \textsl{A stable
three-dimensional topological Dirac semimetal Cd${}_3$As${}_2$}, Nature
Mater. \textbf{13}, 677 (2014).

\bibitem{ZrTe5:2} R. Y. Chen et al.,
\textsl{``Optical spectroscopy study of three dimensional Dirac semimetal
ZrTe$_5$''}, arXiv:1505.00307.

\bibitem{Bi2Se3} Devendra Kumar, Archana Lakhani, \textsl{``Observation of
three-dimensional Dirac semimetal state in topological insulator Bi$_2$Se$_3$%
''}, arXiv:1504.08328.

\bibitem{ref:semimetal:4} L. P. He et al., \textsl{Quantum Transport
Evidence for the Three-Dimensional Dirac Semimetal Phase in Cd${}_3$As${}_2$}%
, Phys. Rev. Lett. \textbf{113}, 246402 (2014) [arXiv:1404.2557].

\bibitem{WeylSemimetalDiscovery} B. Q. Lv et al., \textsl{``Experimental
discovery of Weyl semimetal TaAs''},
arXiv:1502.04684; X. Huang,
\textsl{``Observation of the chiral anomaly induced negative
magneto-resistance in 3D Weyl semi-metal TaAs''}, arXiv:1503.01304; B. Q. Lv
et al.,
\textsl{``Observation of Weyl nodes in TaAs''} arXiv:1503.09188.

\bibitem{VZ} G.E. Volovik and M.A. Zubkov, "Emergent Weyl spinors in
multi-fermion systems," Nuclear Physics B 881, 514 (2014).

\bibitem{W2} A.A. Soluyanov, D. Gresch, Zhijun Wang, QuanSheng Wu, M.
Troyer, Xi Dai, B.A. Bernevig, "Type-II Weyl Semimetals," Nature 527, 495 -
498 (2015).

\bibitem{VolovikBH} P.~Huhtala and G.~E.~Volovik, ``Fermionic microstates
within Painleve-Gullstrand black hole,'' J.\ Exp.\ Theor.\ Phys.\ \textbf{94}
(2002) no.5, 853 [Zh.\ Eksp.\ Teor.\ Fiz.\ \textbf{121} (2002) no.5, 995]
doi:10.1134/1.1484981 [gr-qc/0111055].

\bibitem{VolovikBHW2} G.~E.~Volovik, ``Black hole and Hawking radiation by
type-II Weyl fermions,'' Pisma Zh.\ Eksp.\ Teor.\ Fiz.\ \textbf{104} (2016)
no.9, 660 [JETP Lett.\ \textbf{104} (2016) no.9, 645]
doi:10.7868/S0370274X16210104, 10.1134/S0021364016210050 [arXiv:1610.00521
[cond-mat.other]].

\bibitem{NissinenVolovik2017a} J. Nissinen and G.E. Volovik, Type-III and IV
interacting Weyl points, Pisma ZhETF \textbf{105}, 442--443 (2017) JETP
Lett. \textbf{105}, 447--452 (2017), arXiv:1702.04624.

\bibitem{Volovik2003} G.E. Volovik, \textsl{``The Universe in a Helium
Droplet''}, Clarendon Press, Oxford (2003)

\bibitem{Weinberg} Steven Weinberg, \textsl{''The Quantum Theory of Fields''}%
, Cambridge University Press (1995)

\bibitem{Parrikar2014} Onkar Parrikar, Taylor L. Hughes, and Robert G.
Leigh, \textsl{``Torsion, parity-odd response, and anomalies in topological
states''}, Phys. Rev. D \textbf{90}, 105004 (2014) [arXiv:1407.7043].

\bibitem{Zanelli} O.Chandia and J.Zanelli, \textsl{``Topological invariants,
instantons and chiral anomaly on spaces with torsion,''}\^{A}~ Phys. Rev.D%
\^{A}~55 (1997) 7580\^{A}~ [hep-th/9702025]

\bibitem{Mielke} E.W.Mielke, `\textsl{`Anomalies and gravity,''}\^{A}~ AIP
Conf.Proc.\^{A}~\^{A}~ 857B (2006) 246\^{A}~ [hep-th/0605159]; D.Kreimer and
E.W.Mielke, \textsl{``Comment on: Topological invariants, instantons, and
the chiral anomaly on spaces with torsion,''}\^{A}~ Phys. Rev. D\^{A}~ 63
(2001) 048501\^{A}~ [gr-qc/9904071]

\bibitem{obukhov} Y. Obukhov, \textsl{Spectral geometry of the Riemann -
Cartan space - time}, Nucl.Phys. B212 (1983) 237 - 254

\bibitem{yajima} S. Yajima and T. Kimura, \textsl{Anomalies in
four-dimensional curved space with torsion,} Prog. of Theor. Phys.. 74
(1985) 866 - 880

\bibitem{Obukhov:1997pz} Y.~N.~Obukhov, E.~W.~Mielke, J.~Budczies and
F.~W.~Hehl, \textsl{``On the chiral anomaly in nonRiemannian space-times,''}
Found.\ Phys.\ \textbf{27}, 1221 (1997) [gr-qc/9702011].

\bibitem{soo} C. Soo Phys. Rev. D59, 045006 (1999) hep-th/9805090

\bibitem{semimetal_effects2} A. M. Turner, A. Vishwanath, and C. O. Head,
\textsl{``Beyond band insulators: Topology of semi-metals and interacting
phases''}, Topological Insulators 6 (2013) 293 [arXiv:1301.0330].

\bibitem{semimetal_effects3} G. B. Halsasz and L. Balents, \textsl{%
``Time-reversal invariant realization of the Weyl semimetal phase''}, Phys.
Rev. B \textbf{85}, 035103 (2012).

\bibitem{semimetal_effects6} S. Parameswaran, T. Grover, D. Abanin, D.
Pesin, and A. Vishwanath, \textsl{``Probing the chiral anomaly with nonlocal
transport in Weyl semimetals}, Phys. Rev. X \textbf{4}, 031035 (2014)
[arXiv:1306.1234].

\bibitem{semimetal_effects7} M. N. Chernodub, A. Cortijo, A. G. Grushin, K.
Landsteiner, and M. A. Vozmediano, \textsl{``A condensed matter realization
of the axial magnetic effect''}, Phys. Rev. B \textbf{89}, 081407(R) (2014)
[arXiv:1311.0878].

\bibitem{semimetal_effects8} Z. Jian-Hui, J. Hua, N. Qian, and S. Jun-Ren,
\textsl{``Topological invariants of metals and the related physical effects''%
}, Chinese Phys. Lett. \textbf{30}, 027101 (2013) [arXiv:1211.0772].

\bibitem{semimetal_effects10} M. Vazifeh and M. Franz, \textsl{%
``Electromagnetic response of weyl semimetals''}, Phys. Rev. Lett. \textbf{%
111}, 027201 (2013) [arXiv:1303.5784].

\bibitem{semimetal_effects11} Y. Chen, S. Wu, and A. Burkov, \textsl{``Axion
response in Weyl semimetals''}, Phys. Rev. B \textbf{88}, 125105 (2013)
[arXiv:1306.5344].

\bibitem{semimetal_effects12} Y. Chen, D. Bergman, and A. Burkov, \textsl{%
``Weyl fermions and the anomalous Hall effect in metallic ferromagnets''},
Phys. Rev. B \textbf{88}, 125110 (2013) [arXiv:1305.0183]; David Vanderbilt,
Ivo Souza, and F. D. M. Haldane Phys. Rev. B \textbf{89}, 117101 (2014)
[arXiv:1312.4200].

\bibitem{semimetal_effects13} S. T. Ramamurthy and T. L. Hughes, \textsl{%
``Patterns of electro-magnetic response in topological semi-metals''},
arXiv:1405.7377.

\bibitem{Zyuzin:2012tv} A.~A.~Zyuzin and A.~A.~Burkov, \textsl{``Topological
response in Weyl semimetals and the chiral anomaly,''} Phys.\ Rev.\ B
\textbf{86} (2012) 115133 [arXiv:1206.1868 [cond-mat.mes-hall]].


\bibitem{tewary} Pallab Goswami, Sumanta Tewari, \textsl{Axionic field
theory of (3+1)-dimensional Weyl semi-metals,} Phys. Rev. B 88, 245107
(2013), arXiv:1210.6352


\bibitem{Buividovich2014} P. V. Buividovich, \textsl{''Spontaneous chiral
symmetry breaking and the chiral magnetic effect for interacting Dirac
fermions with chiral imbalance''}, Phys. Rev. D \textbf{90}, 125025 (2014).

\bibitem{SonSpivak2012} D.T. Son and B. Z. Spivak, \textsl{''Chiral anomaly
and classical negative magnetoresistance of Weyl metals''}, arXiv:1206.1627.

\bibitem{SonYamamoto2012} D.T. Son and N. Yamamoto, \textsl{''Berry
curvature, triangle anomalies, and chiral magnetic effect in Fermi liquids''}%
, arXiv:1203.2697.

\bibitem{ZrTe5} Q. Li, D. E. Kharzeev, C. Zhang, Y. Huang, I. Pletikosic, A.
V. Fedorov, R. D. Zhong, J. A. Schneeloch, G. D. Gu, and T. Valla,
arXiv:1412.6543.







\bibitem{Z2015} M.A.Zubkov, \textsl{``Emergent gravity and chiral anomaly in
Dirac semimetals in the presence of dislocations''}, Annals of Phys.,
\textbf{360}, 655 (2015), [arXiv:1501.04998]

\bibitem{Abramchuk:2016afc} R.~A.~Abramchuk and M.~A.~Zubkov, \textit{%
``Schwinger pair creation in Dirac semimetals in the presence of external
magnetic and electric fields,''} Phys.\ Rev.\ D \textbf{94} (2016) no.11,
116012 doi:10.1103/PhysRevD.94.116012 [arXiv:1605.02379 [cond-mat.mes-hall]].

\bibitem{ref:diffusion} Cheng Zhang et al,
\textsl{``Detection of chiral anomaly and valley transport in Dirac
semimetals''}, arXiv:1504.07698.

\bibitem{ref:transport} Tian Liang et al,
\textsl{Ultrahigh mobility and giant magnetoresistance in the Dirac
semimetal Cd${}_3$As${}_2$}, Nature Mater. \textbf{14}, 280 (2015)
[arXiv:1404.7794].

\bibitem{ChiralAnomalySemimetal} Hui Li et al.,
\textsl{``Negative Magnetoresistance in Dirac Semimetal Cd${}_3$As${}_2$''},
arXiv:1507.06470; Cheng Zhang et al.,
\textsl{``Detection of chiral anomaly and valley transport in Dirac
semimetals''}, arXiv:1504.07698; Cai-Zhen Li et al.,
\textsl{``Giant negative magnetoresistance induced by the chiral anomaly in
individual Cd${}_3$As${}_2$ nanowires''}, arXiv:1504.07398; Jun Xiong et
al.,
\textsl{``Signature of the chiral anomaly in a Dirac semimetal: a current
plume steered by a magnetic field''}, arXiv:1503.08179; Jan Behrends, Adolfo
G. Grushin, Teemu Ojanen, Jens H. Bardarson, \textsl{``Visualizing the
chiral anomaly in Dirac and Weyl semimetals with photoemission spectroscopy''%
}, arXiv:1503.04329; Chenglong Zhang et al.,
\textsl{``Observation of the Adler-Bell-Jackiw chiral anomaly in a Weyl
semimetal''}, arXiv:1503.02630; Jun Xiong, Satya Kushwaha, Jason Krizan,
Tian Liang, R. J. Cava, N. P. Ong, \textsl{``Anomalous conductivity tensor
in the Dirac semimetal Na$_3$Bi''}, arXiv:1502.06266.

\bibitem{CVE} J. Erdmenger, M. Haack, M. Kaminski, A. Yarom, JHEP
0901:055,2009

\bibitem{TypeIIWeylSuper} Baruch Rosenstein, B. Ya. Shapiro, Dingping Li, I.
Shapiro, "Magnetic properties of Type I and II Weyl Superconductors",
arXiv:1802.08787

Dingping Li, Baruch Rosenstein, B. Shapiro, I. Shapiro, "Effect of the type
I to type II Weyl semimetal topological transition on superconductivity",
arXiv:1612.09048, 10.1103/PhysRevB.95.094513 Phys. Rev. B 95, 094513 (2017)

\bibitem{ref:LL} L.D. Landau, E.M. Lifshitz, \textsl{``Theory of Elasticity,
Third Edition: Volume 7 (Course of Theoretical Physics)''},
Butterworth-Heinemann, Oxford (1986).

\bibitem{vozmediano2} Fernando de Juan, Juan L. Ma\~nes, Mar\'ia A. H.
Vozmediano, \textsl{``Gauge fields from strain in graphene''}, Phys. Rev. B
\textbf{87}, 165131 (2013) [arXiv:1212.0924].

\bibitem{vozmediano4} M. A. H. Vozmediano, M. I. Katsnelson, F. Guinea,
\textsl{``Gauge fields in graphene''}, Phys. Rep. \textbf{496}, 109 (2010)
[arXiv:1003.5179].

\bibitem{vozmediano5} Alberto Cortijo, Francisco Guinea, Mar\'ia A. H.
Vozmediano, \textsl{``Geometrical and topological aspects of graphene and
related materials''}, J. Phys. A: Math. Theor. \textbf{45}, 383001 (2012)
[arXiv:1112.2054].

\bibitem{vozmediano6} Juan L. Ma\~nes, Fernando de Juan, Mauricio Sturla,
Maria A. H. Vozmediano, \textsl{``Generalized effective Hamiltonian for
graphene under nonuniform strain''}, Phys. Rev. \textbf{88}, 155405 (2013)
[arXiv:1308.1595].

\bibitem{Oliva} M. Oliva-Leyva, G.G. Naumis, \textsl{Understanding electron
behavior in strained graphene as a reciprocal space distortion}, Phys. Rev.
B 88, 085430 (2013), arXiv:1304.6682

\bibitem{VZ2013} G.E. Volovik and M.A. Zubkov, \textsl{``Emergent gravity in
graphene''}, talk presented at the International Moscow Phenomenology
Workshop (July 21-25, 2013), arXiv:1308.2249.

\bibitem{VolovikZubkov2014} G.E. Volovik and M.A. Zubkov, \textsl{``Emergent
Ho{\v r}ava gravity in graphene''}, Ann. Phys. \textbf{340}, 352 (2014)
[arXiv:1305.4665].

\bibitem{Volovik:2014kja} G.~E.~Volovik and M.~A.~Zubkov, \textsl{``Emergent
geometry experienced by fermions in graphene in the presence of
dislocations,'' } Annals Phys. \textbf{356} 255 (2015) [arXiv:1412.2683].

\bibitem{Khaidukov:2016yfi} Z.~V.~Khaidukov and M.~A.~Zubkov, \textit{%
``Landau Levels in graphene in the presence of emergent gravity,''} Eur.\
Phys.\ J.\ B \textbf{89} (2016) no.10, 213 doi:10.1140/epjb/e2016-70182-7
[arXiv:1601.00693 [cond-mat.mes-hall]].

\bibitem{Horava2005} P. Ho\v{r}ava, \textit{"Stability of Fermi surfaces and
$K$-theory"}, Phys. Rev. Lett. \textbf{95}, 016405 (2005).


\bibitem{Chernodub:2015wxa} M.~N.~Chernodub and M.~Zubkov, M.~N.~Chernodub
and M.~Zubkov, \textit{``Chiral anomaly in Dirac semimetals due to
dislocations,''} Phys.\ Rev.\ B \textbf{95} (2017) no.11, 115410
doi:10.1103/PhysRevB.95.115410 [arXiv:1508.03114 [cond-mat.mes-hall]].

\bibitem{Zubkov:2015cba} M.~A.~Zubkov, \textit{``Emergent gravity and chiral
anomaly in Dirac semimetals in the presence of dislocations,''} Annals
Phys.\ \textbf{360} (2015) 655 doi:10.1016/j.aop.2015.05.032
[arXiv:1501.04998 [cond-mat.mes-hall]].

\bibitem{VZ2014NPB} G. E. Volovik, M.~A.~Zubkov, \textsl{``Emergent Weyl
spinors in multi-fermion systems''}, Nucl. Phys. B \textbf{881}, 514 ,\^{a}%
\euro \oe\ (2014) [arXiv:1402.5700].


\bibitem{WeylTightBinding} Shapourian, H., Hughes, T. L., Ryu, S. \textsl{%
The viscoelastic response of topological tight-binding models in 2d and 3d.}
arXiv:1505.03868 (2015)


\bibitem{Cortijo:2015jja} A.~Cortijo and M.~A.~Zubkov, \textit{``Emergent
gravity in the cubic tight-binding model of Weyl semimetal in the presence
of elastic deformations,''} Annals Phys.\ \textbf{366} (2016) 45
doi:10.1016/j.aop.2016.01.006 [arXiv:1508.04462 [cond-mat.mes-hall]].

\bibitem{Z2018} M.~A.~Zubkov, \textit{``The black hole interior and the type
II Weyl fermions,''} Mod.\ Phys.\ Lett.\ A \textbf{33} (2018) no.07n08,
1850047 doi:10.1142/S0217732318500475 [arXiv:1801.00966 [gr-qc]].

\bibitem{conductivity} R. Y. Chen, S. J. Zhang, J. A. Schneeloch, C. Zhang,
Q. Li, G. D. Gu, and N. L. Wang, \textsl{\textquotedblright Optical
spectroscopy study of three dimensional Dirac semimetal ZrTe$_{5}$%
\textquotedblright } arXiv: 1505.00307

\bibitem{crystal_size} A. Pariari, N. Khan, and P. Mandal \textsl{''Magnetic
field induced drastic violation of Wiedemann-Franz law in Dirac semimetal Cd$%
_3$As$_2$''}, ArXiV 1508.02286

\bibitem{crystal_size2} Tian Liang, Quinn Gibson, Mazhar N. Ali, Minhao Liu,
R. J. Cava, N. P. Ong, \textsl{'' Ultrahigh mobility and giant
magnetoresistance in the Dirac semimetal Cd$_3$As$_2$''}, arXiv:1404.7794

\bibitem{Z2017L} M.~A.~Zubkov, \textit{``Fermi points and the Nambu sum rule
in the polar phase of$^{3}$He,''} JETP Lett.\ \textbf{105} (2017) no.11, 721
doi:10.1134/S0021364017110029 [arXiv:1705.02231 [cond-mat.mes-hall]].

\bibitem{LL10} E.M. Lifshitz, L.P.Pitaevsky, \textit{The course of
theoretical Physics,} vol. 10, \textit{Physical Kinetics},
Butterworth-Heinemann, Oxford (1986)

\bibitem{LL9} L.D. Landau, E.M. Lifshitz, \textit{The course of theoretical
Physics,} vol. 9, \textit{Statistical Physical. Part II. Condensed matter
theory.}, Butterworth-Heinemann, Oxford (1986)
\end{thebibliography}
\end{document}